\documentclass{IEEEtran}
\usepackage[utf8]{inputenc}
\usepackage{cite}
\usepackage{amsmath,amssymb,amsfonts}
\usepackage{algorithmic}
\usepackage{graphicx}
\usepackage{textcomp}
\usepackage[caption=false]{subfig}
\usepackage[justification=centering]{caption}
\usepackage[ruled,vlined]{algorithm2e}
\usepackage{epsfig}
\usepackage{float}
\usepackage{color}
\usepackage{balance}
\usepackage{bbm}
\usepackage{textcomp}
\usepackage{enumitem}
\def\BibTeX{\text{B\kern-.05em{\sc i\kern-.025em b}\kern-.08em
    T\kern-.1667em\lower.7ex\hbox{E}\kern-.125emX}}

\newcommand{\E}{{\cal E}}

\newcommand{\beq}{\begin{equation}}
\newcommand{\eeq}{\end{equation}}

\floatstyle{ruled}
\newfloat{algorithm}{tbp}{loa}
\providecommand{\algorithmname}{Algorithm}
\floatname{algorithm}{\protect\algorithmname}

\newtheorem{proposition}{Proposition}
\newtheorem{definition}{Definition}

\newtheorem{remark}{Remark}
\title{\textcolor{black}{Discontinuous Computation Offloading for Energy-Efficient Mobile Edge Computing}}
\author{Mattia Merluzzi,~\IEEEmembership{Member,~IEEE}, Nicola di Pietro, Paolo Di Lorenzo,~\IEEEmembership{Senior Member,~IEEE},\\
Emilio Calvanese Strinati,~\IEEEmembership{Member,~IEEE}, Sergio Barbarossa,~\IEEEmembership{Fellow,~IEEE}\IEEEcompsocitemizethanks{\IEEEcompsocthanksitem M.~Merluzzi and E. Calvanese Strinati are with Univ. Grenoble Alpes, CEA, Leti, F-38000 Grenoble, France, France. \\Email: mattia.merluzzi@cea.fr, emilio.calvanese-strinati@cea.fr.\IEEEcompsocthanksitem N. di Pietro 
is with Athonet, via Cà del Luogo 6/8, 36050, Bolzano Vicentino (VI), Italy. Email: nicola.dipietro@athonet.com. \IEEEcompsocthanksitem P.~Di Lorenzo and S.~Barbarossa are with the Department of Information Engineering, Electronics, and Telecommunications of Sapienza University, via Eudossiana 18, 00184 Roma, Italy.\\E-mail: paolo.dilorenzo@uniroma1.it, sergio.barbarossa@uniroma1.it.
\protect\\
This work was partly supported by the European Commission through the H2020 project Hexa-X (Grant Agreement no. 101015956), by H2020 EU/Taiwan Project 5G CONNI, Nr.~861459, by the CPS4EU project, which has received funding from the ECSEL Joint Undertaking (JU) under grant agreement Nr.~826276, and by MIUR under the PRIN Liquid\_Edge contract.
}} 
\begin{document}
\maketitle
\begin{abstract}
We propose a novel strategy for energy-efficient dynamic computation offloading, in the context of edge-computing-aided beyond 5G networks. The goal is to minimize the energy consumption of the overall system, comprising multiple User Equipment (UE), an access point (AP), and an edge server (ES), under constraints on the end-to-end service delay and the packet error rate performance over the wireless interface. To reduce the energy consumption, we exploit low-power sleep operation modes for the users, the AP and the ES, shifting the edge computing paradigm from an \emph{always on} to an \emph{always available} architecture, capable of guaranteeing an on-demand target service quality with the minimum energy consumption. %
To this aim, we propose an online algorithm for dynamic and optimal orchestration of radio and computational resources called \emph{\textcolor {black}{Discontinuous Computation Offloading} (DisCO)}. In such a framework, end-to-end delay constraints translate into constraints on overall queueing delays, including both the communication and the computation phases of the offloading service. DisCO hinges on Lyapunov stochastic optimization, does not require any prior knowledge on the statistics of the offloading traffic or the radio channels, and satisfies the long-term performance constraints imposed by the users. Several numerical results illustrate the advantages of the proposed method.

\end{abstract}


\section{Introduction}
With the advent of beyond 5G networks \cite{ahmadi20195g,6Gstrinati}, 
mobile communication systems are evolving from a pure communication framework to service enablers, building on the tight integration of communication, computation, caching, and control functionalities  \cite{Barbarossabook2018,Ndikumama19}. Indeed, future networks will 
serve a plethora of new 
applications, not only addressed to mobile end users, but also for whole different sectors (\textit{verticals}), such as Industry 4.0, Internet of Things (IoT), autonomous driving, remote surgery, Artificial Intelligence (AI) etc. These new services have very different requirements and they generally involve massive data processing within low end-to-end (E2E) delays (in the order of ms). Among several technology enablers at different layers (e.g., AI, network function virtualization, millimeter-wave communications), a prominent role will be played by 
Edge Computing, whose aim is to move cloud functionalities (e.g., computing and storage resources) at the edge of the network, to avoid the relatively long delays necessary to reach central clouds. 
Edge Computing is also the object of an ETSI Industry Specification Group, called Multi-Access Edge Computing (MEC) \cite{ETSIMEC}.  
In 5G networks, MEC functionalities will be placed behind the User Plane Function (UPF), thus in the core network or virtualized locally at the Access Point (AP)~\cite{MEC5G}. MEC is foreseen to enable several novel applications and use cases~\cite{MEC_req}, relying on the enhanced performance of new beyond 5G technologies, due to the massive volume of data to be transferred within low-latency and/or extremely high-reliability constraints~\cite{Popovski5G18}. Recent surveys on MEC are available in \cite{Pham2019Survey}, \cite{Mach17}. 

In this paper, we focus on {\it computation offloading} services, in which the execution of applications is transferred from mobile devices (or sensors in IoT environments) to a nearby edge server (ES) \cite{Mach17}. Computation offloading helps reducing the User Equipment's (UE) energy consumption and/or the overall delay of the service. When an application is offloaded, the overall service time is composed of the uplink transmission time of input data, the processing time of this input at the ES, and the time needed to send the results back to the UE~\cite{BarbarossaSardellitti2014,Sardellitti2015}. 
In edge-computing-aided networks, a critical aspect for real-life implementations is the limited energy made available by the battery at the mobile device, the need for frequent battery recharge, and the high energy consumption of network elements, due to the dense deployment of APs and ESs necessary to enable the described ecosystem. In traditional mobile networks, a large portion of the power is consumed at the AP site \cite{AUER11},\cite{Tombaz2015}. With the deployment of ESs, the power consumption will certainly increase, so that new methods are essential to reduce the impact of the ICT industry on the global carbon footprint \cite{Footprint2018}. In such a context, the main target of our paper is the energy efficiency of the overall network, comprising UE, AP, and ES.
\section{Related work and contribution} 
In the context of mobile networks, several works focus on novel strategies to reduce system power consumption.
In general, it is well-known that a large portion of the power is consumed by the AP only for being in active state (RF chains, power amplifiers, cooling, etc.)~\cite{Tombaz2015}. Thus, most of the works in the literature propose strategies based on possible ON/OFF behavior of the APs, known as Discontinuous Transmission (DTX) \cite{Tombaz2015,Bonnefoi18,ChangMiao18,Kim18,Dedomenico12,Dedomenico14,Dedomenico18}, by which some components of the AP are put in low-power sleep states when possible, e.g., in case of low traffic. In the context of edge computing and computation offloading, there exists a wide literature \textcolor{black}{
\cite{Labidi15,Sardellitti2015,You2017,Mao2016,Mao2017,Chen2019,Merluzzi2020URLLC,HanChen2020,ZhouChen2020,Farhanan2020,FangChen2020}}. In particular, \cite{Mao2017} proposes a dynamic computation offloading strategy, based on Lyapunov stochastic optimization, to reduce a weighted sum of UE and ES power consumption. \cite{Chen2019} extends the strategy to a multi-server multi-cell scenario, introducing average delay and reliability constraints on the queue lengths. In \cite{Merluzzi2020URLLC},  a joint dynamic computation offloading strategy was proposed with reliability guarantees, incorporating ultra-reliable low-latency communications and energy harvesting devices. All these works mainly focus on power consumption at the UE and ignore the network. The authors of  \cite{HanChen2020} propose a dynamic strategy aimed at minimizing the average power consumption of mobile devices, under a latency constraint and a constraint on the ES average power consumption, without considering the AP. \textcolor{black}{In \cite{ZhouChen2020}, an auction-based incentive mechanism is proposed to maximize the revenue of a mobile network operator under delay constraints. In \cite{Farhanan2020}, the authors present a multi-objective approach to minimize the execution delay, energy consumption, and monetary cost of the smart devices with service rate constraints. The work in  \cite{FangChen2020} proposes a scheduling and resource provisioning strategy to minimize edge nodes' power consumption under delay and resource constraints.}  Recent contributions consider the energy consumption of both radio access and MEC network~\cite{LiGuan19,ChenZhou17,Wang19,ChangMiao18-2,Nan17,Wu2021,Yu_Pu_2018,Rafia2021}. In particular, in \cite{LiGuan19}, a scheduling strategy is proposed to find a trade-off between task completion ratio and throughput, hinging on Lyapunov optimization, while \cite{ChenZhou17} aims at minimizing the long-term average delay under a long-term average power consumption constraint. In \cite{Wang19}, the long-term average energy consumption of a MEC network is minimized under a delay constraint, using a MEC sleep control. Also, in~\cite{ChangMiao18-2} the problem is formulated as the minimization of the energy consumption under a mean service delay constraint, optimizing the number of active base stations and the computation resource allocation at the ES, while considering a sleep mode for both APs and ESs. In \cite{Nan17}, Lyapunov optimization is used to reduce the energy consumption of a fog network while guaranteeing an average response time. \textcolor{black}{The authors of \cite{Wu2021} minimize the offloading service delay with Lyapunov optimization under constraints on the user's and edge nodes' energy consumption}. In \cite{Yu_Pu_2018}, the authors exploit Lyapunov optimization, Lagrange multiplier, and sub-gradient techniques to optimize devices' and APs' energy consumption under delay constraints, exploiting AP sleep states. \textcolor{black}{The authors of \cite{Rafia2021} propose a method to minimize a weighted sum of users and MEC energy consumption under delay constraints, considering a dedicated time for wireless charging.}

\textcolor{black}{Another class of recent works propose data-driven solutions as, e.g., Deep Reinforcement Learning (DRL) \cite{ChenWang2020,Zhou2021,ZhouWu2021,ZhaoXinjie2020}. In \cite{ChenWang2020}, a decentralized approach based on DRL is proposed to minimize a weighted sum of user local powers, offloading powers, and buffering delays. In \cite{Zhou2021}, the authors solve the problem of computation offloading with a deep Q-network aimed at minimizing the energy consumption of MEC nodes and users under task delay constraints. DRL is also exploited for content caching in \cite{ZhouWu2021}, where the authors aim at maximizing the content provider saving costs, with an incentive mechanism used to motivate end nodes to participate in the offloading process. In \cite{ZhaoXinjie2020}, DRL is used to minimize a weighted sum of energy consumption and delay in an IoT scenario. While \cite{ChenWang2020,Zhou2021,ZhouWu2021,ZhaoXinjie2020} exploit pure data-driven solutions, other recent results show the possibility of merging model-based optimization with the power of data-driven optimization \cite{zapp2019,Bi2021,Bae2020,Sana2021}. 
The common point of all these works is the lack of a \emph{holistic} view of APs' sleep control, radio resource allocation, ESs' sleep and CPU scheduling, and UE's sleep control, under E2E delay constraints, involving average and out-of-service events, which is the goal of this paper. Our main challenge, not addressed in the available literature, is to design a strategy able to deal with complex time-varying scenarios with unknown statistics and several discrete optimization variables, involving heterogeneous entities (i.e. UE, APs and ESs), \textcolor{black}{looking for low-complexity solutions able to run online}.} 
\subsection{Our Contribution}
In this paper, we extend and improve our preliminary results of~\cite{MerluzzidiPietro2019}. In contrast with the state of the art, we \emph{simultaneously} optimize the modulation and coding scheme selection and the power of both the UE's uplink and the AP's downlink radio transmission, the CPU frequency allocation at the ES, and the duty cycles of all the network elements. We propose a dynamic computation offloading strategy based on Lyapunov stochastic optimization that minimizes the weighted sum of UE's, AP's, and ES's long-term average energy consumption, under an \emph{average end-to-end delay constraint} and a \emph{reliability constraint}. The latter is defined as the probability that the end-to-end delay exceeds a prescribed threshold. These constraints are handled through the definition of an uplink queue of data to be offloaded by each UE, a computation queue at the ES, and a downlink queue of results at the AP. These constraints translate into a constraint on the average length of the sum of the three queues and a probabilistic bound on the maximum total queue length, as in \cite{MerluzzidiPietro2019}. However, differently from~\cite{MerluzzidiPietro2019}, we introduce the sleep mode operation at the UE's side and an adaptive algorithm to translate the probabilistic constraint on the queue lengths into a reliability constraint on the \emph{actual} end-to-end delay. Our proposed strategy does not require any \emph{a priori} knowledge of the statistics of the radio channels or of the data arrivals. In particular, starting from a non-convex non-differentiable long-term average optimization problem with unknown statistics, we devise an algorithm that solves a deterministic problem on a per-slot basis, yielding an asymptotically optimal solution of the original problem (as a consequence of Proposition~\ref{proposition}, as explained in Section~\ref{sec:problem_formulation}). The proposed optimal solution of each deterministic problem has very low computational complexity and can be found via 
Algorithm~\ref{alg:radio}, \ref{alg:cpu}, and \ref{alg:DMEC}, presented in Section~\ref{sec:solution_per_slot}. Several numerical results show the performance of our strategy, also compared with other methods, due to the fact that it takes into account the whole network energy consumption, reducing that of all agents \textit{simultaneously}, thus achieving a globally green solution.
\section{System model}\label{sec:system_model}
\begin{figure*}[htb!]
    \centering
    \includegraphics[width=.9\textwidth]{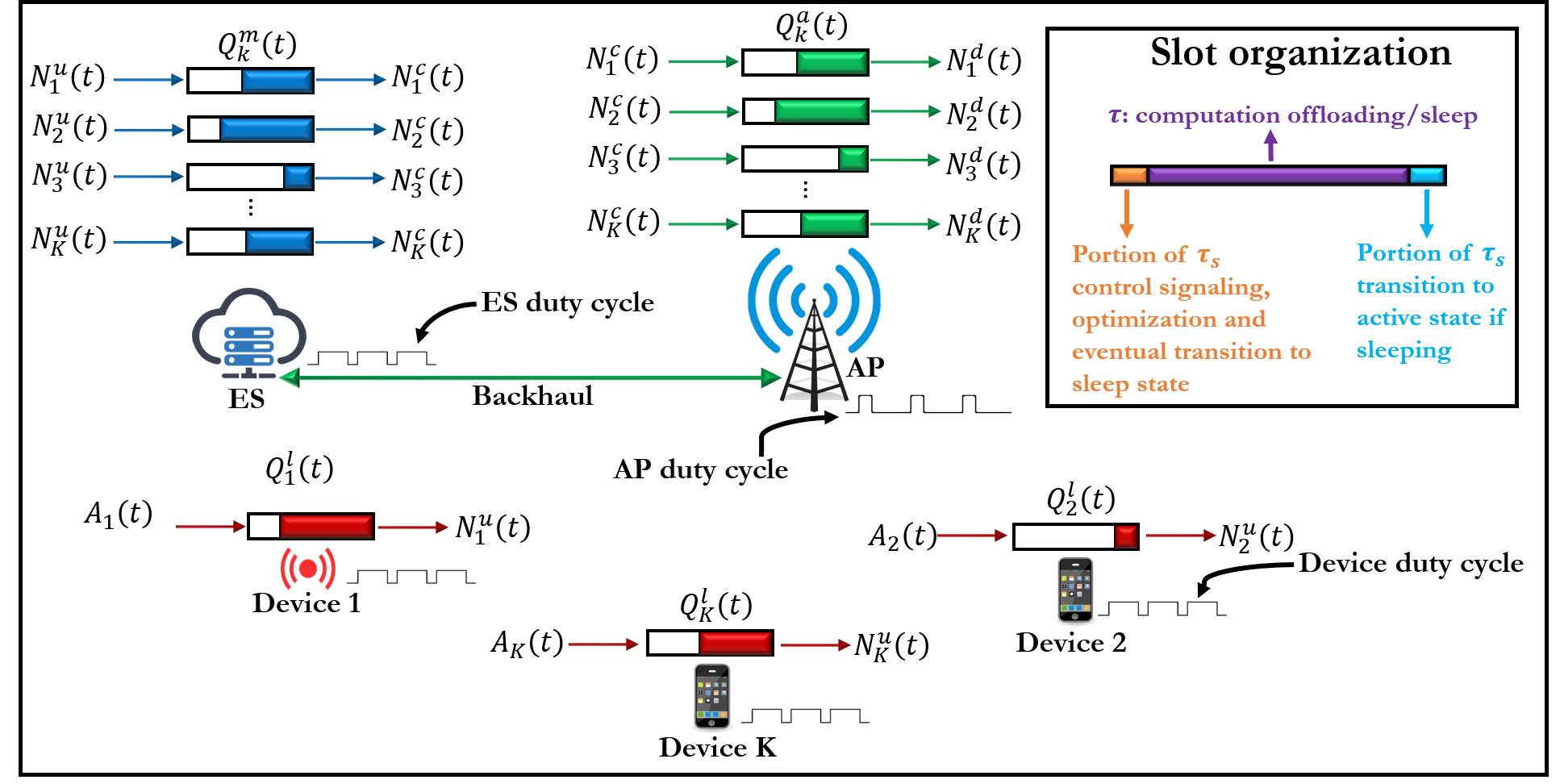}
    \caption{Network model}
    \label{fig:scenario}
\end{figure*}
To capture the dynamic aspects of the problem, we consider time as organized in slots $t=1,2,3\ldots$ of equal duration $\tau_l$. In the following, we present: the UE's, AP's, and ES's energy consumption model; the queueing model used to handle the delay constraints; the reliability performance over the radio interface in terms of Packet Error Rate (PER).
\subsection{Energy consumption model}
Computation offloading generally entails three phases: an uplink phase, where a UE sends data to the AP, a processing phase at the ES, and a downlink phase for the transmission of results to the UE \cite{Mach17,Sardellitti2015}. In our dynamic scenario, the overall slot duration $\tau_l$ is divided into two portions: a period of $\tau_s$ seconds dedicated to control signaling and transition among sleep/active states, and a period of $\tau$ seconds for the actual three phases of computation offloading. 
Indeed, we assume that the AP, the ES, and the UE can enter low-power {\it sleep states} for energy saving purposes during 
the slot fraction reserved to offloading 
(not for the whole slot duration, due to the need for control signaling and state transitions at each time slot). 
When in sleep state, the AP and the UE cannot receive nor transmit and the ES cannot process data, thus consuming less power. Our goal 
is to optimize the long-term fraction of time that the entities spend in sleep state with the aim of minimizing the overall system energy consumption, but guaranteeing a targeted Quality of Service (QoS) measured by the overall delay of the computation offloading service.
Since at the beginning of every slot each network element must be active, before the end of the slot all network elements wake up (if they were sleeping), to be active at the beginning of the next slot for control signaling. Thus, $\tau_s$ comprises a portion at the beginning of the slot, devoted to control signaling and eventual transition from active to sleep, and a portion at the end of the slot, needed to wake up if in sleep state, to be active for control signaling at the beginning of the next slot, as depicted in the top right part of Fig. \ref{fig:scenario}. Finally, the total duration of a time slot is $\tau_l=\tau_s+\tau$. 
It should be noted that, 
when dimensioning $\tau_s$, the transition time has to be taken into account. Thus, in this paper, we exploit sleep operation modes compatible with the slot duration from a transition time point of view, as it will be clarified in the following sections. The transition energy is neglected, as typically done in works related to DTX \cite{Dedomenico12,Dedomenico14,Dedomenico18,Debaille15,Wang19,ChangMiao18-2,Yu_Pu_2018}. However, our model can be easily extended to take into account the transition energy consumption, being just an additional term of power consumed during a sleep phase.
\subsubsection{AP's Energy Consumption}\label{sec:AP_energy}
Nowadays, around 80\% of the total power consumption of the wireless networks is consumed at the AP \cite{Debaille15}, 
which consumes a considerable fraction of its total power only for being in active state~\cite{AUER11,Debaille15}. Let us denote by $p_a^{\rm on}$ the overall power consumption of the AP for being in active state. This parameter generally includes the consumption of power amplifiers, power supply, analog front-end, digital baseband, and digital control. In active state, the AP can transmit and/or receive. Thus, we denote by~$p^d(t)$ the overall downlink transmit power. Let us note that the AP can enter a low-power sleep mode to save energy whenever possible, without compromising 
the QoS. Obviously, 
the deeper the sleep mode, the higher the energy saving, but also the higher the time needed to wake-up, i.e., the minimum sleep period. In Section \ref{sec:num_res}, we will present more specific considerations on the sleep modes, active and sleep power consumption of the AP, and transition times (e.g. from sleep to active). To control the active and sleep state of the AP, we introduce the binary variable $I_a(t) \in \{0,1\}$, which equals $1$ if and only if the AP is in active state at time slot $t$. 
In each slot, the AP is forced to be active for the first portion of $\tau_s$ seconds 
to perform Channel State Information acquisition and control signaling. For simplicity, we neglect the transmit power necessary for this reduced exchange of information, thus taking into account only the active state power $p_a^{\rm on}$ during the signaling period. 
Then, the AP energy consumption at time slot $t$ is 
\begin{equation}\label{AP_energy}
E_a(t)=\tau\left(I_a(t)\left(p_a^{\rm on}+p^d(t)\right)+(1-I_a(t))p_a^{\rm s}\right)+\tau_s p_a^{\textrm{on}},
\end{equation}
where $p_a^{\rm s}$ represents the (low) power consumed by the AP in sleep mode. The power consumed in the receiver chain is neglected, as it is typically much smaller than the other contributions.
\subsubsection{UE's Energy Consumption}\label{sec:UE_energy} 
Going beyond~\cite{MerluzzidiPietro2019}, we assume that all $K$ UE can switch their radio equipment to a low-power sleep mode whenever possible. In particular, we assume that UE $k$ (for $k=1,\ldots,K$) consumes a generic power $p_k^{\rm on}$ only for being active. Also, we denote by $p_k^u(t)$ the power necessary to transmit, assuming that it is a monotone increasing function of the transmit power $p_k^{\rm tx}(t)$. 
In Section \ref{sec:num_res}, we will be more specific with a model from the literature for the typical values of $p_k^{\rm on}$, the function linking $p_k^u(t)$ and $p_k^{\rm tx}(t)$, and the transition times. 
Recalling that the UE is always active at the beginning and the end of the slot for control signaling, the total energy consumption of the UE is
\begin{align}\label{UE_energy}
E_u(t)=\sum_{k=1}^K\bigg[&\tau\left(I_k(t)\left(p_k^{\rm on}+p_k^u(t)\right)+(1-I_k(t))p_k^s\right)\nonumber\\
&+\tau_s p_k^{\rm on}\bigg],
\end{align}
where $I_k(t)$ equals one if UE $k$ is active in time slot $t$, and $0$ otherwise.
\subsubsection{ES's Energy Consumption}\label{sec:ES_energy}
As pointed out in \cite{gough2015energy}, the power management of a CPU is all about efficiently (and dynamically) controlling both current and voltage in order to minimize power while providing a desired performance. Power-saving techniques can be divided into two main categories: \emph{turn it off} and \emph{turn it down}. The first one 
consists in switching off some components of the CPU, which are then put into low-power sleep states. In modern processors, there exist several possible idle states, called C-states \cite{gough2015energy}, which allow the processor to enter more or less deep sleep modes. Obviously, a deeper sleep mode provides higher energy savings, but requires more time to wake up. This defines a trade-off between energy consumption and latency. Furthermore, C-states can operate on each core separately or on the entire CPU package~\cite{brochard2019energy}. 

In this paper, we adopt C-states operating on a specific core, dedicated to treat the offloaded tasks of all the UE of our system. In particular, we consider two states: the C0-state, in which the CPU core is active and executing some thread, and the C$x$-state ($x=1,2,\ldots$), in which the CPU clock frequency is driven to zero. The transition time from C$x$ to C0 depends on the specific choice of the C$x$ state. For instance, for C1-state, it is in the order of $\mu$s \cite{brochard2019energy}. In Section \ref{sec:num_res}, we will present more specific considerations on the choice of the C$x$ state, based on the duration of the slot. In our model, the CPU core consumes a power $p_m^{\rm on}$ just for being in active state (C0-state) and a power $p_m^s$ in sleep state (C$x$-state). Moreover, when the ES is active, the dynamic power spent for computation is 
$p_m^c(t)=\kappa f_c^3(t)$, 
where $f_c(t)$ is the CPU cycle frequency at time slot $t$ and $\kappa$ is the effective switched capacitance of the processor \cite{Burd1996}. We suppose that it is possible to use dynamic voltage frequency scaling to scale down the frequency \cite{LeSueur10}, thus reducing the dynamic power consumption. In particular, we assume that $f_c$ can be selected from a finite set $\mathcal{F}=\{0,\ldots,f_{\max}\}$ 
and we introduce the binary variable $I_m(t)\in\{0,1\}$, which equals $1$ if and only if the ES is in active state. 
Then, recalling that $\tau_l=\tau_s+\tau$, 
the energy consumption in each time slot is given by
\begin{equation}\label{MEH_energy}
E_m(t)=\tau\left(I_m(t)p_m^{\rm on}+ (1-I_m(t))p_m^s+ p_m^c(t)\right)+\tau_s p_m^{\rm on},
\end{equation}
where $I_m(t)=\mathbf{1}\{f_c(t)\}$, with $\mathbf{1}\{\cdot\}$ the indicator function; note that $p_m^c(t) = 0$ whenever $f_c(t) = 0$, because $p_m^c(t)=\kappa f_c^3(t)$. 
Then, from  (\ref{AP_energy}), (\ref{UE_energy}), (\ref{MEH_energy}), the total energy consumption in slot $t$ is:
\begin{equation}\label{total_energy}
E_{\rm tot}(t)=E_u(t)+E_m(t)+E_a(t).
\end{equation}
\subsection{Delay and queueing model}
Computation offloading involves three main steps:
an uplink transmission phase of input data from the UE; a computation phase at the ES; a downlink transmission phase of results back to the UE. We consider a dynamic scenario, in which new input data units are continuously generated from an application at the UE's side and have to be offloaded and processed at the ES. To model the system dynamics, we use a simple queueing model, taking into account the three phases of computation offloading. This model allows us to characterize the total delay experienced by a data unit from its generation at the mobile side until the reception of its corresponding result, sent by the AP to the UE. In particular, the considered queueing model is depicted in Fig.~\ref{fig:scenario}. Specifically, in Fig. \ref{fig:scenario}, we can notice three different queues: $i)$ A local communication queue at each device (red) of data buffered before uplink transmission; $ii)$ A remote computation queue at the ES (blue) of data buffered before being processed; $iii)$ A downlink communication queue (green) of results buffered before being sent back to the devices. Accordingly, each data unit experiences three different delays:
a communication delay, including buffering at the UE; a computation delay, including buffering at the ES; a communication delay, including buffering at the AP.
As we will show later, we take into account these three sources of delay jointly, as in~\cite{Merluzzi2020URLLC}.
For the multiple access over the radio channel, we consider a simple Frequency Division Multiple Access, both for the uplink and the downlink.

\subsubsection{Uplink communication queue}
In uplink, allocating bandwidth $B_k^u$ to UE $k$, the symbol duration is $T_k^{s,u}=\frac{1}{B_k^u}$. Since the time for data transmission is $\tau=\tau_l-\tau_s$, under the assumption of a perfect pulse shaping, UE $k$ can transmit $N_k^{s,u}(t)=\left\lfloor\frac{\tau}{T_k^{s,u}}\right\rfloor=\left\lfloor\tau B_k^u(t)\right\rfloor$ symbols at time $t$. 
Assuming that bits are encoded against radio channel noise into packets of fixed length $N_b$ bits, employing an $M$-QAM modulation, the number of packets transmittable at time $t$ is given by:
\begin{equation}\label{num_pac_up}
    N_k^{p,u}(t)=\left\lfloor\frac{N_k^{s,u}(t)\log_2(M_k^u(t))R_k^{c,u}(t)}{N_b}\right\rfloor
\end{equation}
where $M_k^u(t)$ is the modulation order and $R_k^{c,u}(t)$ is the channel coding rate. In particular, we assume that the uplink Modulation and Coding Scheme (MCS) pair $m_k^u = \left(M_k^{u}(t), R_k^{c,u}(t)\right)$ is chosen from a discrete finite set $\mathcal{M}_k^u$.
Also, we assume that a data unit has to be transferred in one time slot, i.e., it cannot be split and partially transmitted over different time slots. Thus, the number of data units that UE $k$ can send at time slot $t$ over the radio interface is 
\begin{equation}\label{data_up}
    N_k^u(t)=\left\lfloor\frac{N_k^{p,u}(t) N_b}{S_k^i}\right\rfloor,
\end{equation}
where $S_k^i$ is the size in bits of an input data unit. Then, the local queue of input data units to be offloaded evolves as
\begin{equation}\label{q_loc}
Q_k^l(t+1)=\max\left(0, Q_k^l(t)-N_k^u(t)\right)+A_k(t),
\end{equation}
where $A_k(t)$ is the number of newly arrived data units generated by the application running at UE $k$; $A_k(t)$ is modeled as 
a random process whose statistics are not 
known \emph{a priori}.

\subsubsection{Remote computation queue}
We assume that the number of input data units processed by the ES to serve UE $k$ is proportional to the number of CPU cycles allocated for this task. Given the computation rate $f_k(t)$ assigned to user $k$, measured in CPU cycles per second, 
and defining the coefficient $J_k$ as the ratio between the number of processed data and the number of CPU cycles, the number of data units processed by the ES for UE $k$ at time slot $t$ is 
\begin{equation}\label{num_comp}
    N_k^c(t)=\left\lfloor\tau f_k(t)J_k\right\rfloor.
\end{equation}
Hence, the queue of data waiting for being processed by the ES for UE $k$ evolves as
\begin{align}\label{q_rem}
Q_k^m(t+1)=\max(0,Q_k^m(t)-N_k^c(t))+\min(Q_k^l(t),N_k^u(t)).
\end{align}

\subsubsection{Downlink communication queue}
Finally, we define $K$ queues at the AP, containing the computation results to be sent back to the UE. We assume that every processed input data unit produces one output data unit, with size $S_k^o$ possibly different from $S_k^i$. The queue evolves as:
\begin{align}\label{q_ap}
Q_k^a(t+1)=\max\left(0,Q_k^a(t)-N_k^d(t)\right)\!+\!\min(Q_k^m(t),N_k^c(t)),
\end{align}
where $N_k^d(t)$ is the number of output data units sent back to user $k$ in downlink, which is computed as 
\begin{equation}\label{data_down}
    N_k^d(t)=\left\lfloor\frac{N_k^{p,d}(t) N_b}{S_k^o}\right\rfloor,
\end{equation}
where $N_k^{p,d}(t)$ is the number of packets sent in downlink, given similarly as in \eqref{num_pac_up},
using $B_k^d(t)$ for the bandwidth assigned to UE $k$ for downlink communication at time $t$.  $M_k^d(t)$ is the downlink $M$-QAM modulation order, and $R_k^{c,d}(t)$ is the channel coding rate for downlink. As for the uplink, 
the pair $m_k^d = \left(M_k^d(t),R_k^{c,d}(t)\right)$ belongs to a discrete set $\mathcal{M}_k^d$.
 
\subsubsection{End-to-end delay constraints}
As already mentioned, the overall delay experienced by a data unit is the time elapsed from its generation at the mobile side, to the moment the user receives back the result associated with it. By Little's law \cite{little1961}, the average overall service delay is proportional to the average queue length. Then, the overall delay is directly related to the sum of the uplink and downlink communication queues and the computation queue $Q_k^\text{tot}(t)=Q_k^l(t)+Q_k^m(t)+Q_k^a(t)$. 
In particular, given a data unit arrival rate $\displaystyle A_k^{\rm avg}=\mathbb{E}\left\{A_k(t)/\tau_l\right\}$, the long-term average  end-to-end delay experienced by a data unit generated by UE $k$ is $\bar{D}_k^\infty=\lim_{T\to\infty}\frac{1}{T}\sum_{t=1}^T\mathbb{E}\left\{Q_k^{\rm tot}(t)/A_k^{\rm avg}\right\}$, where
the expectation is taken with respect to the random radio channel and data arrival realizations. Our first aim is to guarantee a constraint on the long-term average delay $D_k^{\textrm {avg}}$, written as: 
\begin{equation}\label{avg_q}
\lim_{T\to\infty}\frac{1}{T}\sum_{t=1}^T\mathbb{E}\left\{Q_k^{\rm tot}(t)\right\}\leq Q_k^{\textrm{avg}} = D_k^{\textrm {avg}} A_k^{\rm avg},\quad \forall k.
\end{equation}
As a second objective, we want to ensure a long-term probabilistic constraint on the E2E delay experienced by data units:
\begin{equation}
    \label{eq:delay_prob_constraint}
    \lim_{T\to\infty}\frac{1}{T}\sum_{t=1}^T\textrm{Pr}\left\{D_k(t) > D_k^{\max}\right\}\leq \epsilon_k,\quad \forall k,
\end{equation}
where $D_k^{\max}$ is a predefined threshold, $0 < \epsilon_k < 1$, and $D_k(t)$ represents the overall delay experienced by a generic data unit whose result is received back by UE $k$ at time $t$. The aim of this constraint is to reduce the variability of the delay. 
As mentioned before, there is a direct dependence between the overall delay and the overall queue length, therefore we can translate~\eqref{eq:delay_prob_constraint} into the following probabilistic constraint on the sum of the queues:
\begin{equation}\label{probabilistic}
    \lim_{T\to\infty}\frac{1}{T}\sum_{t=1}^T\textrm{Pr}\left\{Q_k^{\textrm{tot}}(t)>\delta_k Q_k^{\textrm{avg}}\right\}\leq \epsilon_k,\quad \forall k,
\end{equation}
with $\delta_k>1$ conveniently chosen to convert the delay threshold into a queue-length threshold. In principle, there is no direct analytical relation between $\delta_k Q_k^{\textrm{avg}}$ and $D_k^{\max}$, but we will propose in Section~\ref{sec:adaptation_delta} an online method to appropriately select and adapt $\delta_k$. Finally, note that (\ref{probabilistic}) can be equivalently recast as the expectation of a Bernoulli random variable as
\begin{equation}
    \lim_{T\to\infty}\frac{1}{T}\sum_{t=1}^T\mathbb{E}\big\{u\big(Q_k^{\textrm{tot}}(t)-\delta_kQ_k^{\textrm{avg}}\big)\big\}\leq \epsilon_k,\nonumber
\end{equation}
where $u(\cdot)$ is the unitary step function. In the sequel, the event $\{Q_k^{\rm tot}(t)>\delta_k Q_k^{\rm avg}\}$ will be termed as ``\emph{out-of-service}'', and $\epsilon_k$ will be the required \emph{out-of-service probability}.


\subsection{Packet error rate performance}\label{sec:PER}
To satisfy a target performance in terms of packet loss, we want to guarantee that the uplink and downlink PER, denoted respectively ${\rm PER}_k^u$ and ${\rm PER}_k^d$ for UE $k$, do not exceed some targeted thresholds $\theta_k^u$ and $\theta_k^d$. In this sense, given the radio channel state at time $t$, recalling that the transmit power $p_k^{\rm tx}(t)$ used by UE $k$ is a function of the chosen MCS $m_k^u \in \mathcal{M}_k^u$, we define
    $p_k^{\rm tx,min}(m_k^u, t) = \min \left\{ p_k^{\rm tx}(t) : {\rm PER}_k^u\leq \theta_k^u \right\}$. 
A minimum target PER translates into a minimum target SNR $\bar{\gamma}_k$. Thus, the minimum transmit power is 
$p_k^{\rm tx,\min}=\frac{\bar{\gamma}_k N_0 B_k^u}{h_k^u}$, where $N_0$ is the noise power spectral density at the receiver, and $h_k^u$ is the time-varying uplink channel power gain. The same discussion is valid for the downlink transmission. 
\section{Problem formulation 
}\label{sec:problem_formulation}
In this section, we formulate our optimization problem, aimed at minimizing the long-term average weighted sum of the UE's, AP's and ES's energy consumption, \textcolor{black}{as defined in (\ref{AP_energy}),  (\ref{UE_energy}), (\ref{MEH_energy})}, whose value at time slot $t$ is given by the convex combination:
\begin{equation}\label{weigthed_energy}
    E_{\textrm {tot}}^w(t) = \alpha_1 E_u(t)+\alpha_2 E_a(t) + \alpha_3 E_m(t),
\end{equation}
where $\alpha_i\geq0$, $\forall i$, and $\sum_{i=1}^{3}\alpha_i=1$, with the coefficients $\alpha_i$ chosen in order to explore alternative priority mechanisms assigned to different energy consumption sources, as clarified later on. The long-term optimization problem is then:
\begin{align}\label{Problem}
&\underset{\mathbf{\Psi}(t)} \min \;  \displaystyle \lim_{T\to\infty}\frac{1}{T}\sum_{t=1}^{T}{\mathbb{E}\left\{E_{\textrm{tot}}^w(t)\right\}}\smallskip\\
&\hbox{subject to}
\nonumber\\
&\displaystyle (a)\; \text{Eqn.} \;\eqref{avg_q};\nonumber\\
&(b)\;\text{Eqn.} \;\eqref{eq:delay_prob_constraint}; \nonumber\\
&(c)\;\; m_k^u(t)\in\mathcal{M}_k^u, \;\; \forall k,t;\nonumber\\
&(d)\;\; m_k^d(t)\in\mathcal{M}_k^d,\;\; \forall k,t;\nonumber\\
&(e)\;\; p_k^{\rm tx,\min}(m_k^u,t)I_k(t)\leq p_k^{\rm tx}(t)\leq p_k^{\rm tx,\max}I_k(t), \; \forall k,t\nonumber\\
&(f)\;\; I_k(t)\in \{0,1\},\;\forall k,t\nonumber\smallskip;\\
& (g)\;\; p_k^{d,\min}(m_k^d,t)I_a(t)\leq p_k^d(t)\leq p^{d,\max} I_a(t)/K, \; \forall k,t\nonumber\\
&(h)\;\; I_a(t)\in \{0,1\},\;\forall t\smallskip;\nonumber\\
&(i)\; \;\displaystyle f_{\text{c}}(t)\in \mathcal{F}, \; \forall t;\nonumber\\
&(j) \;\displaystyle f_{k}(t)\geq 0, \; \forall k,t;\nonumber\\
&(k)\; \;\displaystyle \sum\nolimits_{k=1}^K f_{k}(t)\leq f_{\text{c}}(t), \; \forall t;\nonumber
\end{align}
where $\mathbf{\Psi}(t)=\left[\{\mathbf{\Phi}_k(t)\}_{k=1}^K,f_c(t),I_a(t)\right]$, with $\mathbf{\Phi}_k(t)=[m_k^u(t),m_k^d(t),p_k^{\rm tx}(t),p_k^d(t),f_k(t),I_k(t)]$.
The constraints in \eqref{Problem} have the following meaning: $(a)$ the average end-to-end delay of each user does not exceed 
$D_k^{\textrm {avg}} =  Q_k^{\textrm{avg}}/ A_k^{\rm avg}$; $(b)$ the out-of-service probability is lower than a threshold $\epsilon_k$; $(c)$-$(d)$ the uplink and downlink MCS belong, respectively, to $\mathcal{M}_k^u$ and $\mathcal{M}_k^d$; $(e)$ the uplink transmit power of each UE guarantee the PER constraints and is lower than some fixed budget $p_k^{\rm tx,\max}$; $(f)$ the indicator variable of each UE's sleep state is binary; $(g)$ the downlink transmit power of each UE guarantee the PER constraints and is lower than some fixed budget $p^{\rm d,\max}/K$; $(h)$ the indicator of the AP's sleep state is binary; $(i)$ the computation frequency of the ES is selected from a discrete set 
$\mathcal{F}$; 
$(j)$ the CPU cycle frequency assigned to UE $k$ is non-negative; $(k)$ the sum of all CPU cycle frequencies assigned to all UE does not exceed the ES's total computation frequency $f_c$.

\textcolor{black}{Clearly, the problem formulation in \eqref{Problem} raises many issues in terms of high complexity and hard tractability. First of all, in \eqref{Problem}, \textcolor{black}{the objective function and the long-term constraints (cf. \eqref{avg_q}, \eqref{eq:delay_prob_constraint}) cannot be computed \textit{a priori}, since the statistics of radio channels and data arrivals are not supposed to be known.} 
Furthermore, even by assuming perfect knowledge of the statistics, several discrete variables over a long-term time horizon are involved, thus making the problem to exhibit exponential computational complexity, in principle. 
Nevertheless, hinging on Lyapunov stochastic optimization \cite{Neely10}, we are able to transform \eqref{Problem} \textit{into a pure stability problem}, which is solved in a per-slot fashion that requires only the observation of instantaneous realizations. \textcolor{black}{Building on stochastic optimization theory, we prove the convergence and the asymptotic optimality of the proposed strategy. Furthermore, we show that the per-slot problem enables a low-complexity solution, even in the presence of the discrete variables, thanks to the decoupling across different slots. Optimality is asymptotically achieved thanks to the introduction of virtual queues that allow the algorithm to keep track online of how well the method is behaving in the real case.} 
In general, different approaches can be followed when the system model (or part of it) is not known or to handle complexity efficiently. For instance, 
in \cite{Zhou2021}, the authors approximate the original long-term problem, which is a mixed integer linear program that exhibits exponential complexity, using a Markov decision process that is then solved via DRL. The main difference of our work with respect to the data-driven solution of \cite{Zhou2021} is that, by exploiting the mathematical models presented in Section \ref{sec:system_model} \textcolor{black}{and keeping track of the instantaneous (real and virtual) queues' state, it is possible to split the original problem into a series of consecutive simpler problems that do not need a reinforcement method to be solved, but rather enjoy closed form expressions and fast iterative solutions, with asymptotic theoretical guarantees.}}
\subsection{Lyapunov stochastic optimization}
We present now a way to guarantee the long-term constraints, based on Lyapunov  stochastic optimization. The solution depends on the definition of two \textit{virtual queues} for each UE. For each UE $k=1,\ldots K$, \textcolor{black}{the} first virtual queue $Z_k(t)$, used to impose constraint $(a)$ in \eqref{Problem}, evolves as
\begin{equation}\label{virtual_Z}
    Z_k(t+1)=\max\left(0,\ Z_k(t) + Q_k^{\textrm{tot}}(t+1)-Q_k^{\textrm{avg}}\right).
\end{equation}
Similarly, for constraint $(b)$, we define a virtual queue $Y_k(t)$ that evolves as 
\begin{align}\label{virtual_Y}
Y_k(t+1) =& \max (0,\ Y_k(t)\nonumber\\
&+\mu_k\left(u \left\{Q_k^{\textrm{tot}}(t+1) - \delta_k Q_k^{\rm avg}\right\}-\epsilon_k\right)),
\end{align}
where $\mu_k>0$ is a scalar stepsize.
 The \emph{mean rate stability} of the queues is defined as~\cite[p. $17$]{Neely10}:
\begin{equation}\label{mean_rate}
    \lim_{T\to\infty}\frac{\mathbb{E}\{Z_k(T)\}}{T}=0,\quad\forall k,\quad \lim_{T\to\infty}\frac{\mathbb{E}\{Y_k(T)\}}{T}=0,\quad \forall k.
\end{equation}
In particular, the mean-rate stability of $Z_k(t)$ and $Y_k(t)$ is sufficient to ensure constraints $(a)$ and $(b)$ in \eqref{Problem} \cite{Neely10}. 
We now introduce the \emph{Lyapunov function} \cite[p.~$32$]{Neely10}: $$L(\mathbf{\Theta}(t))=\frac{1}{2}\sum\nolimits_{k=1}^K\left[Z_k^2(t)+Y_k^2(t)\right],$$ 
where $\mathbf{\Theta}(t)=[\{Z_k(t)\}_k,\{Y_k(t)\}_k]$. From 
$L(\mathbf{\Theta}(t))$, we can define the \textit{conditonal Lyapunov drift}~\cite[p.~$33$]{Neely10}, which is the conditional expected variation of $L(\mathbf{\Theta}(t))$ over one slot
\begin{equation}\label{cond_drift}
\Delta(\mathbf{\Theta}(t))=\mathbb{E}\{L(\mathbf{\Theta}(t+1))-L(\mathbf{\Theta}(t))|\mathbf{\Theta}(t)\}.
\end{equation}
Minimizing \eqref{cond_drift} is enough to achieve~\eqref{mean_rate}, but may yield the drawback of 
an unnecessary energy consumption. For this reason, we need to integrate the objective function of \eqref{Problem} in the drift, obtaining the \emph{drift-plus-penalty function} \cite[p.~$39$]{Neely10}:
\begin{equation}\label{drift_plus_penalty}
    \Delta_p(\mathbf{\Theta}(t))=\Delta(\mathbf{\Theta}(t))+ V\cdot\mathbb{E}\{E_{\textrm{tot}}^w(t)|\mathbf{\Theta}(t)\},
\end{equation}
where $V$ is a trade-off parameter used to tune the relative importance given to the objective function with respect to the average virtual queue backlog. 
$\Delta_p(\mathbf{\Theta}(t))$ 
is just a ``penalized'' version of \eqref{cond_drift}. Then, the parameter $V$ is used to trade off the average weighted energy consumption in \eqref{weigthed_energy} and the average E2E delay, as it will be also clarified in Section \ref{sec:num_res} with the numerical results. Now, proceeding as in \cite{Neely10}, we minimize an upper bound of \eqref{drift_plus_penalty} in each time slot, whose derivation is described in the appendix (cf. \eqref{upper_bound}). In particular, our method requires the solution of the following optimization problem in each time slot:
\begin{align}\label{slot_opt}
    &\underset{\mathbf{\Psi}(t)} \min\;\; \sum_{k=1}^K\bigg[-2Q_k^l(t)N_k^u(t) + 4Q_k^m(t) \left(N_k^u(t) - N_k^c(t) \right) \nonumber\\
    &+ 4Q_k^a(t)\!\left(N_k^c(t)- N_k^d(t) \right)\!+\!Z_k(t)\big[\max\left(0,Q_k^l(t)-N_k^u(t)\right)\nonumber\\
    &+\max\left(0,Q_k^m(t)-N_k^c(t)\right)+\max\left(0,Q_k^a(t)-N_k^d(t)\right)\big]\nonumber\\
    &+\mu_k Y_k(t)u\big\{\max\left(0,Q_k^l(t)-N_k^u(t)\right)+ A_k(t) \nonumber\\
    &+ \max\left(0,Q_k^m(t)-N_k^c(t)\right)+\min\left(Q_k^l(t),N_{k,\max}^u\right)\nonumber\\
    &+\max\left(0,Q_k^a(t)-N_k^d(t)\right)+ \min\left(Q_k^m,N_{k,\max}^c\right) \nonumber\\
    &- \delta_k Q_k^{\rm avg}\big\}\bigg] + V E_{\textrm{tot}}^w(t)\nonumber\\
    & \displaystyle \hbox{\text{subject to}} \quad \mathbf{\Psi}(t)\in \mathcal{Z}(t),
\end{align}
where $\mathcal{Z}(t)$ is the feasible set 
definde by $(c)$-$(k)$ of \eqref{Problem}. 
Now, at every 
$t$, the \textit{Min Drift-Plus-Penalty Algorithm} observes the queue states $Q_k^l(t)$, $Q_k^m(t)$, $Q_k^a(t)$, $\mathbf{\Theta}(t)$ and the random events $h_{k}(t)$, $A_k(t)$ and produces a control decision $\mathbf{\Psi}(t)\in\mathcal{Z}(t)$ based on the solution of \eqref{slot_opt}. The non-convex non-differentiable objective function in \eqref{slot_opt} is difficult to optimize. 
Thus, 
we proceed by finding a suitable approximation of \eqref{slot_opt} that simplifies the solution but still provides optimality guarantees. In particular, we hinge on the concept of \textit{$\Gamma$-additive approximation} \cite[p. $59$]{Neely10}: 

\begin{definition}
For a given constant $\Gamma$, a $\Gamma$-\textit{additive approximation}  of the drift-plus-penalty algorithm is one that, for a given state  $\mathbf{\Theta}(t)$ at slot $t$, chooses a (possibly randomized) action $\mathbf{\Psi}(t)\in \mathcal{Z}(t)$ that yields a conditional expected value of the objective function in \eqref{slot_opt} that is within a constant $\Gamma$ from the infimum over all possible control actions. 
\end{definition}



To find a suitable $\Gamma$-approximation, we first introduce the following upper bound, used to get rid of the non-linearity introduced by the $\lfloor\cdot\rfloor$ operator in \eqref{slot_opt}. In particular, since we can write $x-1 \leq \lfloor x \rfloor \leq x$, we have
    $\max(0,Q_k^m(t)-\lfloor\tau f_k(t)J_k\rfloor)\leq\max(0,Q_k^m(t)-\tau f_k(t)J_k+1)$. 
Then, adding without loss of generality the following additional constraint:
\begin{equation}
    f_k(t)\leq \min\left(\frac{Q_k^m(t)+1}{\tau J_k},f_c\right),\quad \forall k,t,\nonumber
\end{equation}
we have 
    $\max(0,Q_k^m(t)-\tau f_k(t)J_k+1)=Q_k^m(t)-\tau f_k(t)J_k+1$. 
Finally, to deal with the non-linearity introduced by the step function
$u\{\cdot\}$, we note that 
$
    u\{x-A\}\leq u\{x\}\leq x+1, \ \forall x,A\geq 0$.
Applying these bounds to the objective function of~\eqref{slot_opt} and removing the constant terms,
the problem can be re-formulated as follows (we omit the temporal index $t$ for ease of notation):
\begin{align}\label{slot_opt2}
    &\underset{\mathbf{\Psi}} \min\quad \sum_{k=1}^K\bigg[(4Q_k^m-2Q_k^l) N_k^u-\widetilde{Q}_k\tau f_kJ_k-4Q_k^aN_k^d\nonumber\\
    &\qquad\qquad\quad+(Z_k+\mu_k Y_k)\big(\max(0,Q_k^l-N_k^u)\nonumber\\
    &\qquad\qquad\quad+\max(0,Q_k^a-N_k^d)\big)\bigg]+VE_{\textrm{tot}}^w
\displaystyle \hspace{1.5cm}\nonumber\\
&\hbox{\text{subject to}}
\nonumber\\
&(a)\quad \mathbf{\Psi}\in \mathcal{Z};\quad (b)\quad \;\displaystyle 
f_{k}\leq \min\left(\frac{Q_k^m+1}{\tau J_k},\ f_c\right), \; \forall k;
\end{align}
where 
$\widetilde{Q}_k=4(Q_k^m-Q_k^a)+Z_k+\mu_k Y_k$
and $\mathbf{\Psi}$ and $\mathcal{Z}$ are defined as for~\eqref{slot_opt}. The following theoretical result applies.

\begin{proposition}\label{proposition}
Suppose that the channel gains $\{h_k(t)\}_k$ and the data arrivals $\{A_k(t)\}_k$ are i.i.d over time, that \eqref{Problem} is feasible, and that $\mathbb{E}\{L(\mathbf{\Theta}(0))\}<\infty$; then, solving \eqref{slot_opt2} in each time slot guarantees that all virtual queues are mean-rate stable (i.e., \eqref{mean_rate} holds) and $E_{\rm tot}^{w}$ is such that:
\begin{equation}\label{Optimality}
\limsup\limits_{T\rightarrow\infty}\frac{1}{T}\sum_{t=1}^{T}\mathbb{E}\{E_{\rm tot}^w(t)\}\leq E_{\rm tot}^{w,\rm opt} + \frac{\zeta+\Gamma}{V},
\end{equation}
where $E_{\rm tot}^{w,\rm opt}$ is the infimum time average energy achievable by any policy that meets the required constraints, and $\zeta$ is a positive constant defined in the appendix (cf. \eqref{zeta}).
\end{proposition}

\begin{IEEEproof}
The proof follows from the fact that the control policy deriving from the solution of \eqref{slot_opt2} is a $\Gamma$-additive approximation of the drift-plus-penalty algorithm in \eqref{slot_opt}. This holds true because, for any given state  $\mathbf{\Theta}(t)$ of the physical and virtual queues at slot $t$, function \eqref{slot_opt} is bounded from above due to the finite size of the feasible set $\mathcal{Z}(t)$, for all $t$. Thus, the  conditional expected value of the objective function in \eqref{slot_opt}, evaluated in the solution of \eqref{slot_opt2}, is within a constant $\Gamma$ from the global optimum of problem \eqref{slot_opt}. The derivations leading to \eqref{slot_opt} and \eqref{zeta} are given in the Appendix. The main claim comes as a direct consequence of \cite[Th. 4.8]{Neely10}.
\end{IEEEproof}

\begin{remark}
\emph{The main consequence of Proposition~\ref{proposition} is that the mathematically optimal long-term solution of~\eqref{Problem} is achieved by optimally solving~\eqref{slot_opt2}, as $V$ increases. 
This will be also clearly visible in the numerical results of Section~\ref{sec:num_res}. 
}
\end{remark}

\begin{remark}
\label{rem:two_pbs}
\emph{The algebraic manipulations that led to~\eqref{slot_opt2}, 
decouple the radio and computation optimization variables and
allow us to 
optimally split the main problem into two different sub-problems: i) radio resource allocation problem, both in uplink and downlink; ii) ES CPU scheduling problem. 
}

\end{remark}
We now present a low-complexity algorithm 
that achieves the globally optimal solution of~\eqref{slot_opt2}.
\section{Solution of the per-slot optimization problem}
\label{sec:solution_per_slot}

\subsection{Radio Resource Allocation}
\noindent The problem for radio resource allocation involves: (i) the decision on the UE and AP sleep state, (ii) the uplink transmit power and MCS selection, and (iii) the downlink transmit power and MCS selection. Then, omitting the temporal index $t$, defining $\mathbf{\Gamma}=[\{m_k^u\}_{k},\{m_k^d\}_{k},\{p_k^{\rm tx}\}_{k},\{p_k^d\}_{k},\{I_k\}_{k},I_a],$ and recalling~\eqref{UE_energy}, \eqref{total_energy}, and~\eqref{weigthed_energy}, the first sub-problem reads as:
\begin{align}\label{slot_opt_radio}
    &\underset{\mathbf{\Gamma}} \min\, \sum_{k=1}^K\bigg[(4Q_k^m-2Q_k^l) N_k^u-4Q_k^aN_k^d\nonumber\\
    &+(Z_k+\mu_k Y_k)\big(\max(0,Q_k^l-N_k^u)+\max(0,Q_k^a-N_k^d)\big)\nonumber\\
    &+V\alpha_1 E_k\bigg]+V\alpha_2 E_a\qquad \displaystyle \nonumber\\
    &\hbox{\text{subject to}}
\quad \mathbf{\Gamma}\in \mathcal{Z}',
\end{align}
where $\mathcal{Z}'$ is the feasible set for the radio resources according to $(c)$-$(h)$ of \eqref{Problem}. Now, to solve~\eqref{slot_opt_radio}, we can distinguish between two different cases:\\
\textbf{Case 1: $I_a=0$}. In this case, since the AP is in sleep mode and cannot receive nor transmit, no user can transmit or receive and $I_k=0$, for all $k$. Thus, the trivial solution is $N_k^u=N_k^d=p_k^{\rm tx}=p_k^d=0$. Moreover, recalling that each UE and the AP are forced to be in active state for a period $\tau_s$ necessary for control signaling, the energy consumption of each user is simply given by $E_k=\tau p_k^s+\tau_s p_k^{\rm on}$, while the energy consumption of the AP is $E_a=\tau p_a^s+\tau_s p_a^{\rm on}$. Then, the minimum value of the objective function of \eqref{slot_opt_radio} in this case, is given by:
\begin{align}\label{obj_sleep}
L^s = &\sum_{k=1}^K(Z_k+\mu_k Y_k)(Q_k^l+Q_k^a)\nonumber\\
&+ V\bigg[\alpha_1\sum_{k=1}^K(\tau p_k^s+\tau_s p_k^{\rm on})+\alpha_2(\tau p_a^s+\tau_s p_a^{\rm on})\bigg].
\end{align}
This value will be compared with the solution obtained in the following second case.\\
\textbf{Case $2$: $I_a=1$}. In this case, the AP is available for transmission and/or reception, and the radio resources in uplink and downlink can take different values. In particular each UE can optimize its $I_k$. Thus, we can distinguish between the case $I_k=0$, in which no transmission or reception occurs for UE $k$, and the case $I_k=1$, in which the uplink and the downlink resources can take any value of the feasible set. When $I_k=0$, we have $N_k^u = N_k^d = p_k^{\rm tx}=p_k^{d}=0$ and the part of the objective function associated with each UE is
\begin{equation}\label{user_sleep}
    L_k^s=(Z_k+\mu_k Y_k)(Q_k^l+Q_k^a)+V\alpha_1\big(\tau p_k^s + \tau_s p_k^{\rm on}\big).
\end{equation}
On the other hand, in the case $I_k=1$, the optimization of each uplink and downlink variable is independent from the others.
We now show the solutions for each user in the case $I_k=1$.

\subsubsection{Optimal Uplink Radio Resource Allocation}
As already mentioned, in this work we assume that the spectral resources (i.e., the bandwidth) are assigned \emph{a priori}. 
This makes the problem separable among different UE and can be formulated, for 
all $k$ with $I_k=1$, as follows:
\begin{align}\label{slot_opt_power_up}
&\underset{\{m_k^u,p_{k}^{\rm tx}\}} \min   \;\displaystyle (4Q_k^m-2Q_k^l)N_k^u+(Z_k+\mu_k Y_k)\max(0,Q_k^l-N_k^u)\nonumber\\
&+ V\alpha_1\tau p_k^u\ +\  V\alpha_1(\tau + \tau_s)p_k^{\rm on}\nonumber \\
&\displaystyle \hspace{.3cm} \hbox{\text{subject to}}\nonumber\\
&
(a)\quad m_k^u\in\mathcal{M}_k^u;\quad (b)\quad p_k^{\rm tx,\min}(m_k^u)\leq p_{k}^{\rm tx}\leq p_k^{\rm tx,\max},
\end{align}
where we recall that $p_k^u$ is a (given) monotone increasing function of $p_k^{\rm tx}$ (cf.~Section~\ref{sec:UE_energy}, and $N_k^u$ is a function of $m_k^u$ (cf. \eqref{data_up}). Since $\mathcal{M}_k^u$ is discrete and finite, \eqref{slot_opt_power_up} can be easily solved via an exhaustive search over all possible schemes in $\mathcal{M}_k^u$ (with linear complexity in the cardinality of $\mathcal{M}_k^u$), where the optimal choice for the transmit power for each $k$ is
    $p_k^{\rm tx} = p_k^{\rm tx,\min}(m_k^u)$. 
Note that it might happen that $p_k^{\rm tx,\min}(m_k^u)\geq p_k^{\rm tx,\max}$. In this case, the selected MCS cannot be used to guarantee the required PER; thus, the solution of~\eqref{slot_opt_power_up} has to be searched in the subset of~$\mathcal{M}_k^u$ that satisfies the constraint on the PER. We denote by $m_k^{u,\rm opt}$ and $p_k^{\rm tx, opt}$ the optimal values of the MCS and the uplink transmit power, respectively. Then, the optimal value of $N_k^u$ is obtained by plugging $m_k^{u, \rm opt}$ in \eqref{data_up}.
Finally, If no MCS can be used to guarantee the required PER, the user $k$ does not transmit, i.e., $N_k^{u, \rm opt} = p_k^{\rm tx, opt}=0$.

\subsubsection{Optimal Downlink Radio Resource Allocation}
The downlink resource allocation is similar to the uplink case, so that the following subproblem of~\eqref{slot_opt_radio} is solved for each user in each slot:
\begin{align} 
\label{slot_opt_power_down}
&\underset{\{m_k^d,p_k^d\}} \min   \displaystyle -4Q_k^a N_k^d + (Z_k\!+\!\mu_k Y_k)\max(0,Q_k^a-N_k^d)\!+\!V\alpha_2p_k^d\nonumber \\
&\displaystyle \hspace{.3cm} \hbox{\text{subject to}}\nonumber\\
&(a)\quad m_k^d\in\mathcal{M}_k^d; \quad  (b)\quad p_k^{d,\min}(m_k^d)\leq p_{k}^d\leq p^{d,\max}/K, 
\end{align}
where $N_k^d$ is a function of $m_k^d$ (cf. \eqref{data_down}). The solution of this problem is obtained, as for the uplink case,  via an exhaustive search over the feasible values of $m_k^d$. We denote by $m_k^{d,\rm opt}$ and $p_k^{d,\rm opt}$ the optimal solutions of \eqref{slot_opt_power_down}. The optimal value of $N_k^d$, denoted by $N_k^{d,\rm opt}$ is obtained by plugging $m_k^{d, \rm opt}$ into \eqref{data_down}. Then, let $L_k^a$ be the following quantity, resulting from the UE's active state:
\begin{align}\label{obj_ue_active}
    &L_k^a=(4Q_k^m-2Q_k^l) N_k^{u,\rm opt}-4Q_k^a N_k^{d,\rm opt}\nonumber\\
    &+(Z_k+\mu_kY_k)\big(\max(0,Q_k^l-N_k^{u,\rm opt})\nonumber\\
    &+\max(0,Q_k^a-N_k^{d,\rm opt})\big)\nonumber\\
    &+V\alpha_1\big(\tau p_k^{u,\rm opt}+(\tau+\tau_s)p_k^{\rm on}\big)+V\alpha_2\tau p_k^{d,\rm opt}.
\end{align}
The optimal value of $I_k$, denoted by $I_k^{\rm opt}$, is then chosen based on the comparison between \eqref{user_sleep} and \eqref{obj_ue_active}. In particular, $I_k^{\rm opt}=1$ if $L_k^a<L_k^s$, and $I_k^{\rm opt}=0$, otherwise. 
\begin{algorithm}[t]
\SetAlgoLined
 In each time slot $t$, observe $Q_k^l, Q_k^m, Q_k^a, Z_k, Y_k, h_k^u, h_k^d$, $\forall k$.\\
\textbf{S1.} Solve \eqref{slot_opt_power_up} and \eqref{slot_opt_power_down} to find for each UE $k$ the values $m_k^{u,\rm opt}, p_k^{\rm tx, opt}, m_k^{d,\rm opt}, p_k^{d,\rm opt}$. Plug $m_k^{u,\rm opt}$ and $m_k^{d,\rm opt}$ into \eqref{data_up} and \eqref{data_down} to find $N_k^{u,\rm opt}$ and $N_k^{d,\rm opt}$, respectively.\\
\textbf{S2.} Compute $L_k^s$ and $L_k^a$ from \eqref{user_sleep} and \eqref{obj_ue_active}, respectively, $\forall k$. \\
  \textbf{S3.} \For{$k=1,\ldots, K$}{ \eIf{$L_k^s\leq L_k^a$}{$I_k^{\rm opt}=0,\quad N_k^{d,\rm opt}=N_k^{u,\rm opt}=p_k^{d,\rm opt}=p_k^{\rm tx, opt}=0$.}{$I_k^{\rm opt}=1$.}
   }
  \textbf{S4.} Compute $L^s$ and $L^a$ from \eqref{obj_sleep} and \eqref{active_state}, respectively. \\
  \textbf{S5.} \eIf{$L^s\leq L^a$}{
  
   $I_a^{\rm opt}=0,\quad N_k^{d,\rm opt} = N_k^{u,\rm opt} = p_k^{d, \rm opt} = p_k^{\rm tx, opt} = 0$, $\forall k$. 
   }{
   $I_a^{\rm opt}=1$.
  }
\caption{Radio Resource Allocation}
\label{alg:radio}
\end{algorithm}
Finally, letting
\begin{equation}\label{active_state}
    L^a=\sum_{k=1}^K\left(I_k^{\rm opt}L_k^a + (1-I_k^{\rm opt})L_k^s\right)+V\alpha_2(\tau+\tau_s)p_a^{\rm on},
\end{equation}
the optimal value of $I_a$, denoted by $I_a^{\rm opt}$, is chosen based on the comparison between \eqref{obj_sleep} and \eqref{active_state}. In particular, $I_a^{\rm opt}=1$ if $L^a<L^s$, and $I_a^{\rm opt}=0$, otherwise. The overall procedure for the optimal radio resource allocation in uplink and downlink is summarized in Algorithm \ref{alg:radio}.
\subsection{Optimal CPU scheduling}
The sub-problem of \eqref{slot_opt_radio} for CPU scheduling at the ES is formulated as follows:
\begin{align}\label{slot_opt_cpu}
    &\underset{\mathbf{\Phi}} \min\ \  V\alpha_3\tau\bigg(I_m(p_m^{\rm on}-p_m^s) + p_m^s+ \kappa f_c^3\bigg)\nonumber\\
    &\qquad-\tau \sum_{k=1}^K\widetilde{Q}_k f_k J_k + V\alpha_3\tau_s p_m^s  \\
&\displaystyle \hspace{.5cm} \hbox{\text{subject to}}\nonumber\\
&(a)\; f_c\in\mathcal{F};\quad(b)\; 0\leq f_k\leq \min\left(\frac{Q_k^m+1}{\tau J_k},f_c\right),\quad \forall k;\nonumber\\
&(c)\;\sum_{k=1}^K f_k\leq f_c,\nonumber
\end{align}
with $\mathbf{\Phi}=[f_c,\{f_k\}_k]$, and we recall that $I_m=\mathbf{1}\{f_c\}$. From (\ref{slot_opt_cpu}), we notice that, for a fixed $f_c$, the problem is linear in the variables $\{f_k\}_k$ and can be efficiently solved via a fast iterative algorithm. Thus, we can perform a search for the optimal value of $f_c$ within $\mathcal{F}$. In particular, the overall procedure to select the optimal $f_c$, the ES's sleep variable $I_m$, and the optimal scheduling frequencies $\{f_k\}_k$ is described in Algorithm \ref{alg:cpu}. Steps S$2$-S$6$ find the optimal CPU resource allocation for a given $f_c$: to minimize $L_c$, we need to allocate the maximum possible CPU frequency to the UE with the highest $\widetilde{Q}_k$; if this leaves some available CPU frequency (cf. step S$3$), the same principle is applied to the remaining UE. Note also that the $|\mathcal{F}|$ iterations over the possible $f_c$ (steps S$1$-S$7$) can be easily parallelized, being independent from each other. From a complexity point of view, even when not parallelized, it is important to notice that Algorithm \ref{alg:cpu} requires, in the worst case, $K\times|\mathcal{F}|$ iterations. 
\begin{algorithm}[t]
\SetAlgoLined
 In each time slot $t$, observe $Q_k^m$, $Q_k^a$, $Z_k$, $Y_k$.\\
 Define the $|\mathcal{F}|\times 1$ vector of the available CPU frequencies $\varphi=\mathcal{F}$. Define the $|\mathcal{F}| \times K$ matrix $F=\{F_{ik}\}_{i,k}$, and the $|\mathcal{F}|\times 1$ vector $L= \{L_i\}_{i=1}^{|\mathcal{F}|}$. Set $F_{ik} = 0\ \forall i,k$, and $L_i=0$ $\forall i$.\\
 \For{$\rm i=1,\ldots,|\mathcal{F}|$}{
\textbf{S1.} Let $\tilde{\varphi} = \varphi_i$, $I_m = \mathcal{I}\{\tilde{\varphi}\}$, and  $\mathcal{U}=\{k=1,\ldots,K\}$.\\
\While{$\tilde{\varphi}>0$ and $\mathcal{U}\neq\emptyset$}{
  \textbf{S2.}  $\displaystyle \tilde{k}=\arg \max_{k\in \mathcal{U}}\left\{J_k\widetilde{Q}_k\right\}$.\\
  \textbf{S3.} $\displaystyle F_{i \tilde{k}} = \min\left((Q_{\tilde{k}}^m+1)/(\tau J_{\tilde{k}}),\ \tilde{\varphi}\right)$.\\
\textbf{S4.} $\mathcal{U}=\mathcal{U} \smallsetminus \left\{\tilde{k}\right\}$.\\
  \textbf{S5.} $\tilde{\varphi} = \tilde{\varphi} - F_{i \tilde{k}}$.
%
   }
   \textbf{S6.} Define $\bar{k}=\{k:\widetilde{Q}_k\leq 0\}$, and set $f_{\bar{k}}=0$.\\
   \textbf{S7.} Compute the objective function $L_c$ of \eqref{slot_opt_cpu} with $f_c= \varphi_i$ and $f_k = F_{ik}$, $\forall k$; save it in $L_i$.
   }
   \textbf{S8.} Find $\displaystyle i^*=\arg \min_{i}\{L_i\}$, $f_c^{\rm opt}= \varphi_{i^*},
   \, f_k^{\rm opt} = F_{i^* k}\; \forall k.$ 
\caption{ES CPU Scheduling}
\label{alg:cpu}
\end{algorithm}
\subsection{End-to-end probabilistic delay constraint adaptation}\label{sec:adaptation_delta}
We now present an online adaptation method to set the parameter $\delta_k$ so that~\eqref{probabilistic} accurately represents~\eqref{eq:delay_prob_constraint}. 
Given a starting point $\delta_k(0)$, the parameter is updated at each time slot as follows:
\begin{equation}\label{adaptation_delta}
    \delta_k(t)=\max \left(\delta_k(t-1)-\nu_k(t)(P_k(D_k^{\max},W_k^t)-\epsilon_k),\ 1 \right),
\end{equation}
where $P_k(D_k^{\max},W_k(t))$ is a moving estimate of the out-of-service probability evaluated on the set $W_k^t$ (of size $|W_k^t|$) composed by the last data units received by UE $k$ until $t$: 
\begin{equation}\label{probs}
    P_k(D_k^{\max},W_k^t)=\frac{1}{|W_k^t|}\sum_{w\in W_k^t}\mathcal{I}\{D_k^w>D_k^{\max}\},
\end{equation}
where $D_k^w$ is the end-to-end delay of the $w$-th data unit in $W_k^t$. In particular, $|W_k(t)|$ is the minimum between a given value (chosen to accurately estimate the probability), and the actual number of received data until time $t$, due to the fact that, at the beginning, there might be no sufficient data to estimate the probability. Furthermore, $\nu_k(t)$ is a stepsize sequence, typically chosen either constant or using the diminishing rule
\begin{equation}\label{diminishing}
    \nu_k(t)=\frac{\nu_k(0)}{t^{\beta_k}}, \quad \beta_k\in (0,1].
\end{equation}
The rationale behind the adaptation rule in \eqref{adaptation_delta} is the following: first, we know from the theoretical analysis of 
Section~\ref{sec:problem_formulation} and from Proposition~\ref{proposition} that, for a given $\delta_k$, Algorithm~\ref{alg:radio} and~\ref{alg:cpu} yield a solution to~\eqref{Problem} that satisfies~\eqref{probabilistic}. Therefore, if~\eqref{probabilistic} is satisfied but the current estimated $P_k(D_k^{\max},W_k^t)$ is actually greater than the desired value $\epsilon_k$, it means that $\delta_k Q_k^{\rm avg}$ misrepresents $D_k^{\max}$ and it is actually greater than it should. Consequently, $\delta_k$ has to decrease at the next time step in order to impose a tighter threshold and let $\delta_k Q_k^{\rm avg}$ better represent $D_k^{\max}$. The opposite happens instead, when $P_k(D_k^{\max},W_k^t) < \epsilon_k$, to achieve a lower-energy solution. Finally, the overall dynamic strategy is described in Algorithm \ref{alg:DMEC}, which will be termed as DisCO.


\begin{algorithm}[t]
\SetAlgoLined
\textbf{Input data}: $K$, $\mathcal{F}$, $B_k^u$, $B_k^d$, $p_k^{\rm tx,\max}$, $p^{d,\max}$, $J_k$, 
$\mathcal{M}_k^u$, $\mathcal{M}_k^d$, $\alpha_1$, $\alpha_2$, $\alpha_3$, $D_k^{\rm avg}$, $D_k^{\max}$, $\epsilon_k$, $\mu_k$.\\
In each slot $t$ do:\\
\textbf{S1.} Find the optimal radio and computation resource allocation with Algorithms \ref{alg:radio} and \ref{alg:cpu}, respectively, and run accordingly the computation offloading procedure.\\
\textbf{S2.} Update the physical and virtual queues as in \eqref{q_loc}, \eqref{q_rem}, \eqref{q_ap}, \eqref{virtual_Z} and \eqref{virtual_Y}, respectively.\\
\textbf{S3.} Update $\nu_k$ as in \eqref{diminishing}, estimate $P_k(D_k^{\max},W_k^t)$ as in \eqref{probs}, and update $\delta_k$ as in \eqref{adaptation_delta}.\vspace{0.2 cm}

\caption{Discontinuous Computation Offloading (DisCO)}
\label{alg:DMEC}
\end{algorithm}
\section{Numerical results}\label{sec:num_res}
\begin{figure*}[t]
    \centering
    \subfloat[Avg. delay vs. weighted system energy]{
        \includegraphics[width=0.32\textwidth]{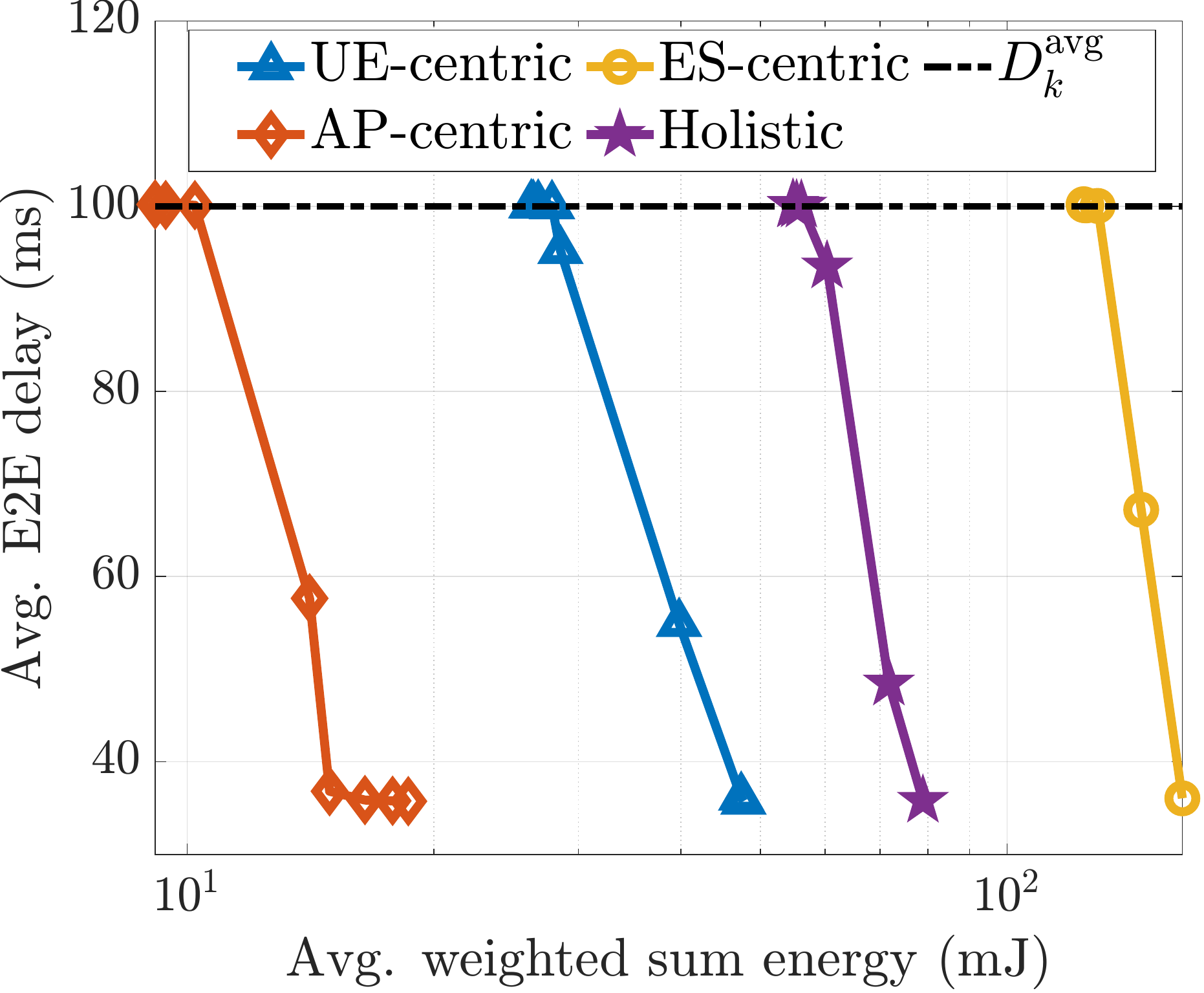}
        \label{fig:tradeoff}
    }
    \subfloat[Average UE energy vs. $V$]{
        \includegraphics[width=0.32\textwidth]{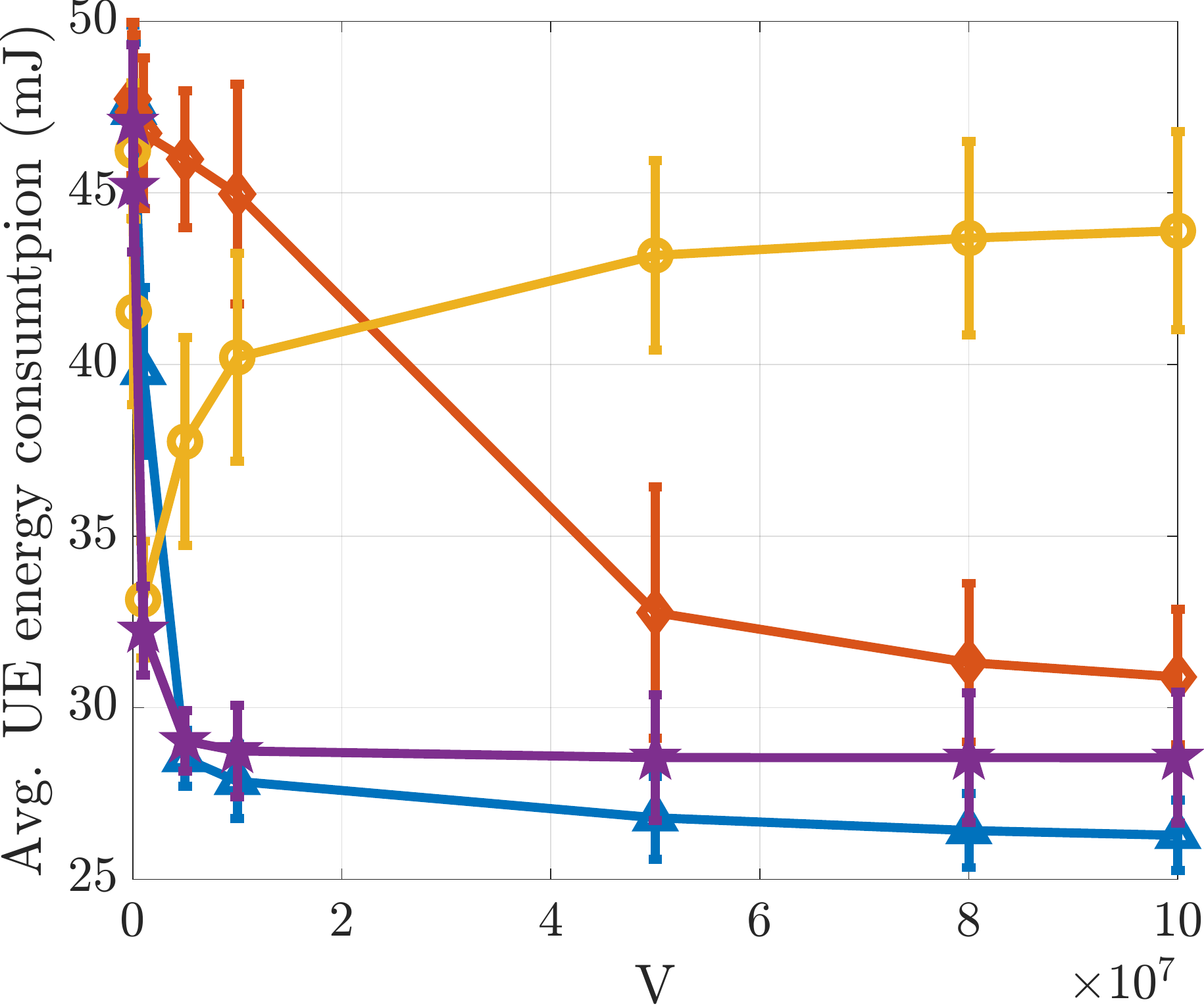}
        \label{fig:UE_energy_vs_V}
    }
    \subfloat[Average AP energy vs. $V$]{
        \includegraphics[width=0.32\textwidth]{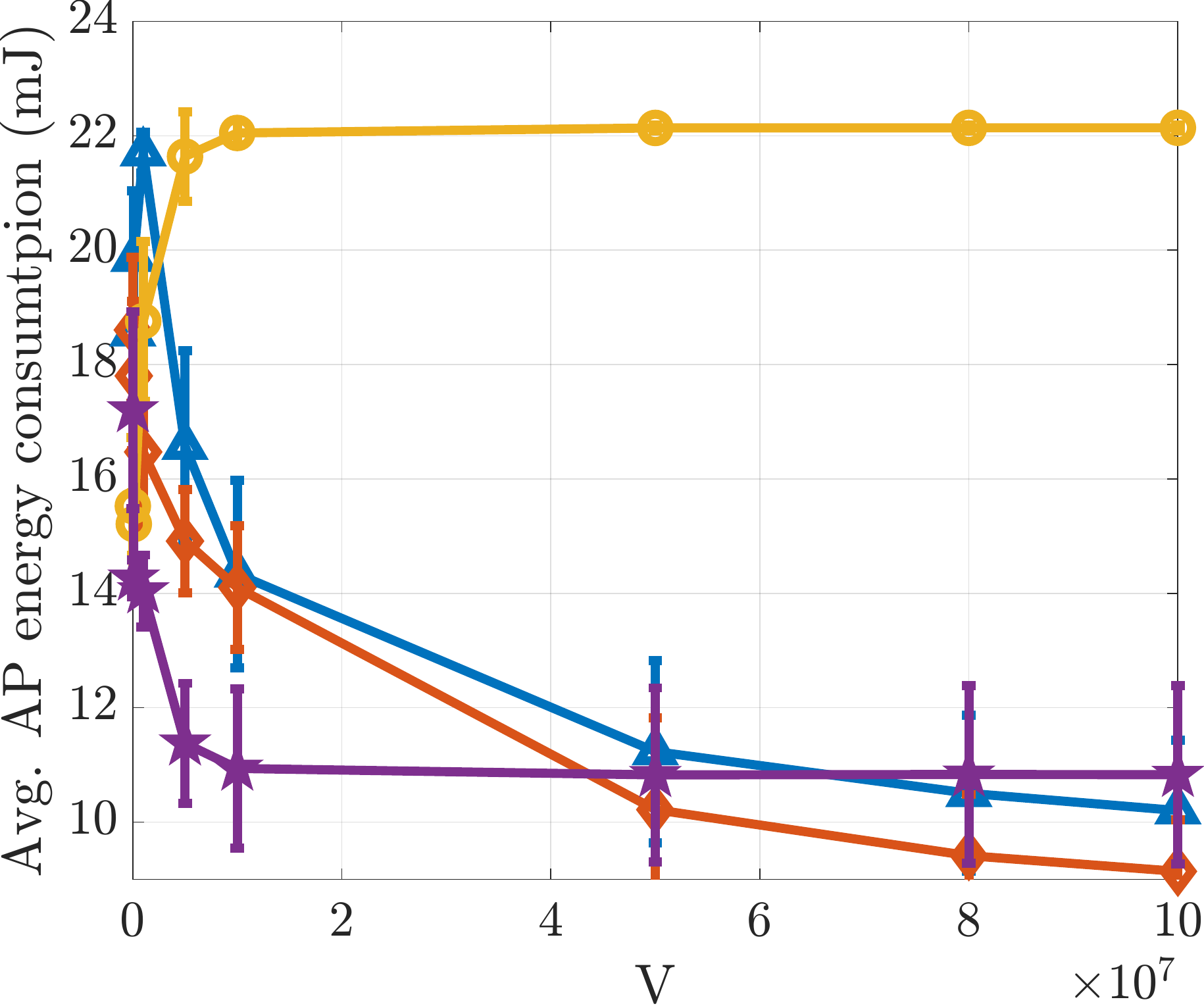}
        \label{fig:AP_energy_vs_V}
    }
    
    \subfloat[Average ES energy vs. $V$]{
        \includegraphics[width=0.32\textwidth]{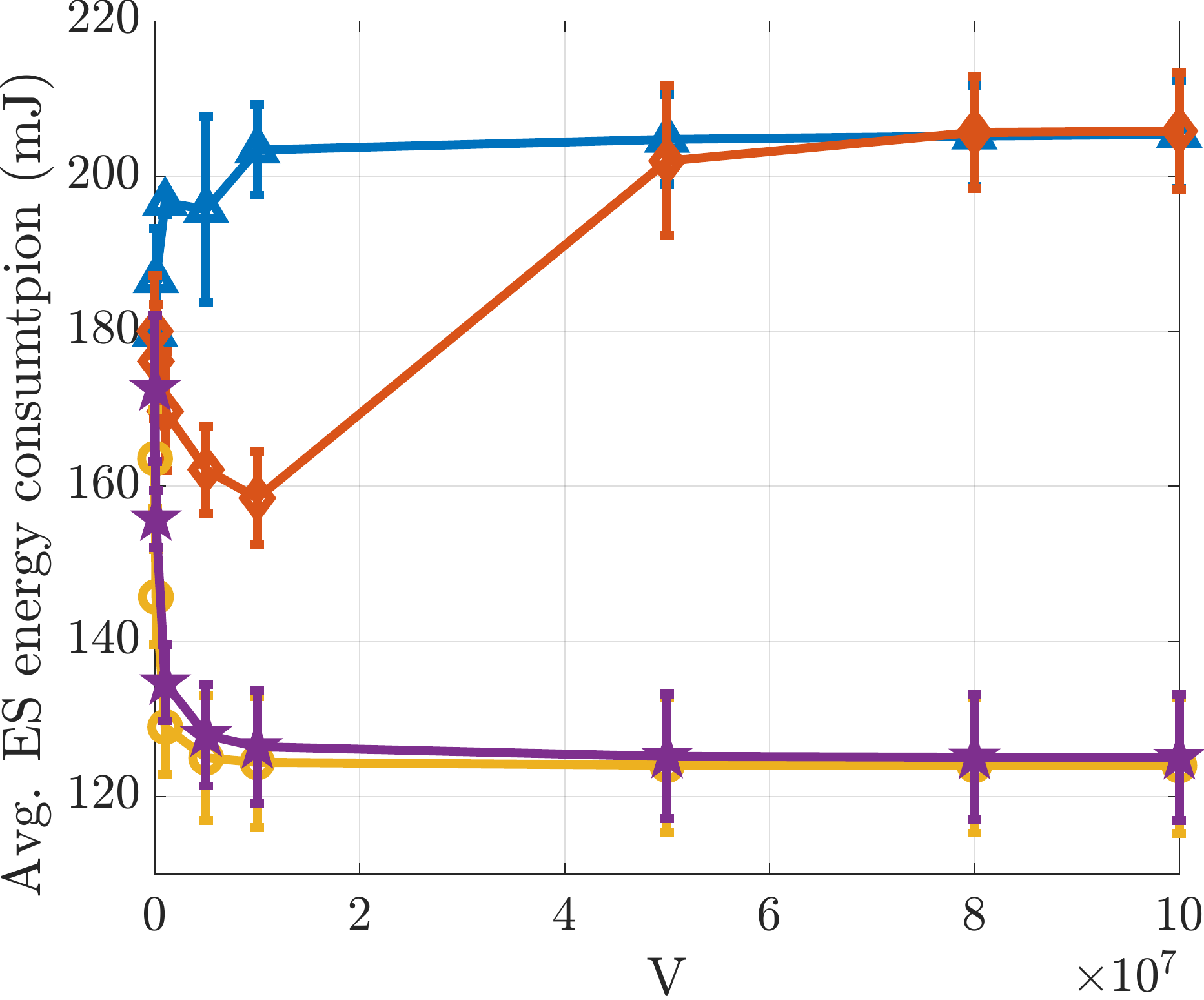}
        \label{fig:server_energy_vs_V}
    }
    \subfloat[Average system energy vs. $V$]{
        \includegraphics[width=0.32\textwidth]{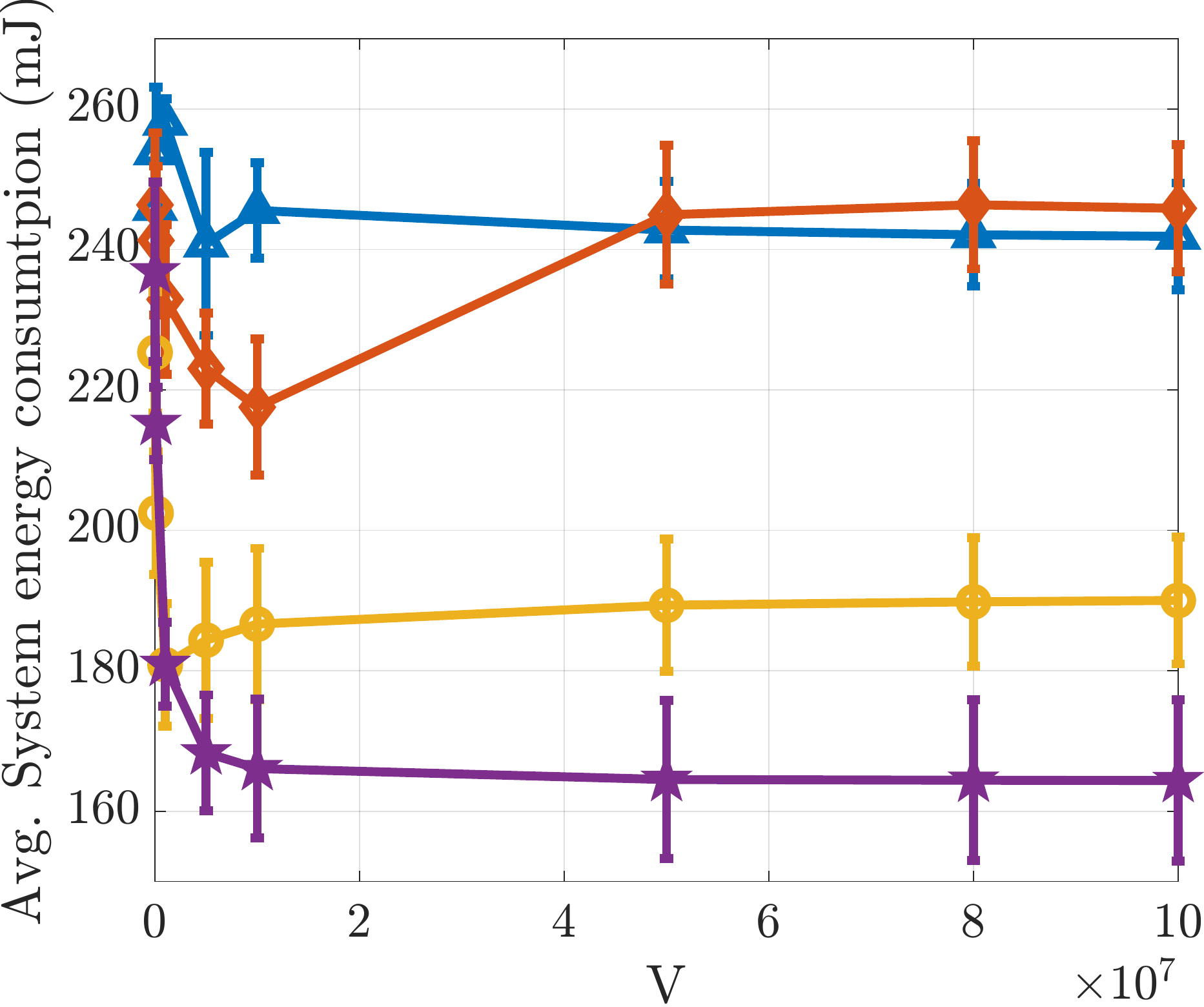}
        \label{fig:system_energy_vs_V}
    }
    \caption{Energy-delay trade-off.
    }
    \label{fig:energy_delay}
\end{figure*}
We present here simulation results to assess the performance of our online optimization strategy. All simulations are performed in Matlab\textregistered,  with the following fixed settings.\\
\textbf{Fixed settings.} We consider a picocell placed at the center of a square area of side $150$ m. We assume an FDD system, with total available bandwidth $B=10$ MHz equally split between uplink and downlink. We consider a total time slot duration $\tau_l=10$ ms, with $\tau_s=1$ ms the portion of the slot used for control signaling and optimization, i.e. where all entities are in active state. Then, the slot duration for data transmission and computation is $\tau=9$ ms. Therefore, the sleep modes of all entities have to be selected according to these values, taking into account the specific transition times. For the AP power consumption, among the different models available in the literature, we exploit that of~\cite{Debaille15}, which provides a tool, available online, to model the power consumption of base stations of different kinds, with details on the specific components (power amplifiers, supply power, etc.). However, our proposed optimization strategy is not constrained to the use of this model; it is more general and can be applied to different models. Recalling the notation of Section \ref{sec:AP_energy}, in the case of a picocell, the AP active power is $p_a^{\rm on}=2.2$~W. The maximum transmit power of the AP is set to $251$~mW \cite{IMECmodel}, 
so that the maximum transmit power of each user is $
251/K$ mW. In \cite{Debaille15}, four possible Sleep Modes (SM) are defined, with different minimum sleep periods, corresponding to the OFDM symbol, the sub-frame duration, the radio frame duration, and a standby mode. For our simulations, we exploit Sleep Mode $2$ from \cite{Debaille15}, whose minimum sleep time is $1$ ms, while a single transition (e.g. from sleep to active) requires $0.5$ ms \cite{Debaille15}, so that a total duration $\tau_s=1$ ms is enough, considering $0.5$ ms at the beginning of the slot (for optimization and eventual transition to sleep state) and $0.5$ ms at the end of the slot to wake up and being active at the beginning of the next slot (see up right part of Fig. \ref{fig:scenario}). The power consumption in sleep mode $2$ 
is $p_a^s=278$~mW \cite{Debaille15}.

The channel model is taken from \cite{SunRap16}, with a carrier frequency of $
28$ GHz, and a Rayleigh fading with unit variance. The noise power spectral density is $N_0=-174$ dBm/Hz, with an additional noise figure of $5$ dB both at UE and at the AP. For the UE power consumption, recalling the notation of Section \ref{sec:UE_energy}, we exploit the empirical model of \cite{Lauridsen14}, where it is shown that the active power is about 
$p_k^{\rm on}=0.9$~W, and is also affected by transmit powers above $10$~mW, consuming an additional $0.6$-$1.5$ W. Here, we assume a maximum transmit power $p_k^{\rm tx,\max}=100$~mW per UE. According to~\cite{Lauridsen14}, the power $p_k^u(t)$ consumed to transmit is a monotone increasing function of the transmit power $p_k^{\rm tx}(t)$. 
For the sleep operation, similarly to the AP case, two different states are defined~\cite{Lauridsen14}: a light sleep mode, with power consumption $p_k^s=346$ mW and sub-ms transition time, and a deep sleep mode, with $p_k^s=20.3$ mW and much longer transition time (around $10$ ms). In this paper, we exploit the light sleep operation, due to the sub-ms transition time. For the numerical model presented in Section~\ref{sec:PER}, we can choose all $M$-QAM modulations with $M \in \{4,16,64,256\}$, coupled with coding rates in $\{0.3,0.4,0.5,0.6,0.7,0.8,0.9\}$, both in uplink and in downlink, so that $\mathcal{M}_k^u$ and $\mathcal{M}_k^d$ have 
$28$ elements. The packet length used in Section~\ref{sec:PER} is $1500$ bytes. The ES has a maximum CPU cycle frequency $f_{\max}=4.5\times 10^9$ CPU cycle/s and an effective switched processor capacitance $\kappa=10^{-27}$ $\text{W}\cdot\left(\frac{\text{s}}{\text{CPU cycle}}\right)^3$~\cite{Burd1996}. The vector of all possible CPU cycle frequencies is 
$\varphi=[0,0.1,0.2,0.3,0.4,0.5,0.6,0.7,0.8,0.9,1]\times f_{\max}$. Finally, recalling the notation of Section \ref{sec:ES_energy}, the power consumption in active state is 
$p_m^{\rm on}=20$ W, whereas the sleep state power consumption is $p_m^s=10$ W.\\
\underline{\textit{Energy-Delay trade-off:}} As a first numerical result, we illustrate the performance of DisCO in terms of energy-delay trade-off. We run our simulations with random configurations of the following parameters: the input and output data size $S_k^i=10^x$, $S_k^o=10^y$ bits, with $x$ and $y$ uniformly randomly generated (u.r.g.) in $[2,3]$ and $[1,3]$, respectively; we assume Poisson arrivals with $A_k^{\rm avg}$ u.r.g. in $[5,15]$ data units; finally, $J_k=10^{-z}$ data/CPU cycle, with $z$ u.r.g. in $[2,5]$. The simulation has run for $T=10^5$ slots and it has been repeated over $100$ independent realizations of the above random parameters and of $K=5$ users' positions, uniformly distributed in a square of side $150$~m. All UE have an average delay requirement $D_k^{\rm avg}=100$~ms and, for this simulation, $\delta_k$ is fixed for all $k$, with $\delta_k = [1.5,1.6,1.7,1.8,1.9]$. The out-of-service constraint is $\epsilon_k=10^{-2}$, with stepsize 
$\mu_k=10$, $\forall k$ (cf.~\eqref{virtual_Y}). We assume the bandwidth to be equally shared among all UE and a target PER of $10^{-4}$, both in uplink and downlink. In Fig. \ref{fig:tradeoff}, we show the trade-off between the long-term average of \eqref{weigthed_energy} and the average E2E delay, defined in \eqref{avg_q}. This trade-off is obtained by increasing the Lyapunov parameter $V$ (cf. \eqref{drift_plus_penalty}) from right to left, as shown in the figure. We plot this trade-off for different settings of the weighting parameters $\alpha_i$, $i=1,2,3$ in \eqref{weigthed_energy}, which also correspond to proper customization of works previously appeared in the literature to our framework and system model. $i)$ \textbf{UE-centric setting ($\blacktriangle$):}  This strategy is obtained by setting $\alpha_1=1$, $\alpha_2=\alpha_3=0$ (cf. \eqref{weigthed_energy}), to only consider the UE energy consumption. This strategy could be possibly related to our previous work \cite{Merluzzi2020URLLC}, where only the UE's energy consumption is optimized.
$ii)$ \textbf{AP-centric setting ($\blacklozenge$):} This strategy is obtained by setting $\alpha_2=1$, $\alpha_1=\alpha_3=0$, to consider only the AP energy consumption. A radio -centric optimization is proposed in \cite{Yu_Pu_2018}, where the authors aim to minimize the sum of UE's and AP's energy consumption in a multi-AP scenario.
$iii)$ \textbf{ES-centric setting ($\bullet$):} This strategy is obtained by setting $\alpha_3=1$, $\alpha_1=\alpha_2=0$, to only consider the ES energy consumption;
$iv)$ \textbf{Holistic solution ($\bigstar$):} This strategy is obtained by setting $\alpha_1=\alpha_2=\alpha_3=1/3$, to take into account the \textit{overall network energy consumption}.
We can notice how the average weighted energy decreased as $V$ increases, while the average E2E delay increases until reaching its maximum value $D_k^{\rm avg}$ imposed by constraint $(a)$ of \eqref{Problem}, as suggested by the theoretical result in Proposition \ref{proposition}, for all strategies. Looking at Fig. \ref{fig:tradeoff}, one may conclude that the AP-centric strategy ($\blacklozenge$) is the best one because it achieves the best trade off between average weighted sum energy and delay. However, this does not give any clue on the energy consumption of the single agents and the network. Therefore, we now wonder what is the behavior of the single sources of energy consumption. Let us notice that, for the highest values of $V$, all strategies reach the same E2E delay, so that we can compare them in terms of energy consumption, given an E2E delay. Thus, in Fig. \ref{fig:energy_delay}b-e, we show the long-term average energy consumption of all users, the AP, the ES, and the overall energy consumption (the sum of the three), all as a function of the Lyapunov tradeoff parameter $V$ (cf.~\eqref{drift_plus_penalty}), with the same value of $V$ as for Fig. \ref{fig:tradeoff}.  Some comments follow:  
\begin{enumerate}[label=(\alph*)]
\item \textbf{UE-centric setting ($\blacktriangle$)}. In this setting, 
the energy consumption of the UE (Fig. \ref{fig:UE_energy_vs_V}) reaches its lowest level, while the energy consumption of the ES (Fig. \ref{fig:server_energy_vs_V}) is not optimized. Instead, 
the AP's energy consumption (Fig. \ref{fig:AP_energy_vs_V}) reaches a level very close to its lowest, obtained with the AP-centric setting ($\blacklozenge$). This is due to the fact that the AP tends to operate in sleep mode when no UE transmits or requests results back, which happens often in the user-centric setting. 

\item \textbf{AP-centric setting ($\blacklozenge$).} In this case, 
the energy consumption of both the AP and the UE approach lower values, for similar reasons as the previous case. This suggests that there exists a strong link between the two entities, since they must be active at the same time when they need to 
communicate. We can interpret (a) and (b) as ``radio-centric'' solutions. 

\item \textbf{ES-centric setting ($\bullet$).} This solution, 
yields the lowest possible energy consumption for the ES as expected, but it is detrimental for the radio part, incurring additional energy consumption for the AP and the users. 
\item \textbf{Holistic solution ($\bigstar$).} This solution aims at minimizing the overall system energy consumption, 
This is the most interesting and promising strategy, since it is globally ``green'' 
and it reaches very close-to-optimal energy consumption for each agent (UE, AP, ES). This suggests that the three sources of energy consumption can be minimized jointly without detrimental effects on the single agents. Practically, the choice of the $\alpha_i$ 
is based on the particular needs of the telecom operator, the MEC operator, or the UE, but could be also based on a global and holistic energy reduction policy. In this paper, we do not tackle the problem of optimizing the $\alpha_i$ for the different needs and leave it for future investigation. 
\end{enumerate}
This first result motivates us to fix 
$\alpha_i=1/3$, $i=1,2,3$ (holistic solution) for the next simulations.\\
\begin{figure*}[t]
    \centering
        \subfloat[Reliability function]{
    \includegraphics[width=0.33\textwidth]{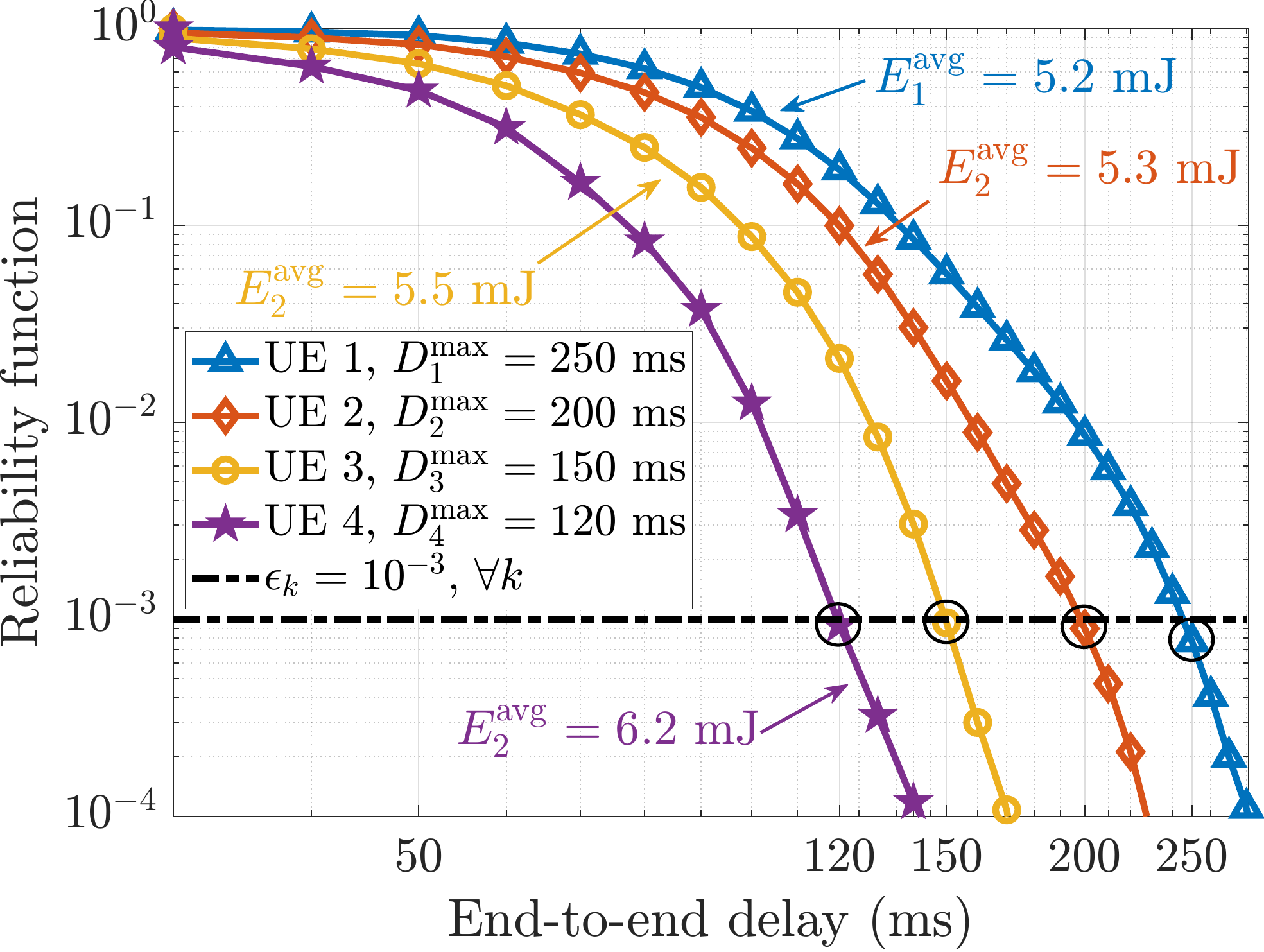}
    \label{fig:reliability}
    }
    \subfloat[Adaptation of $\delta_k$]{
    \includegraphics[width=0.32\textwidth]{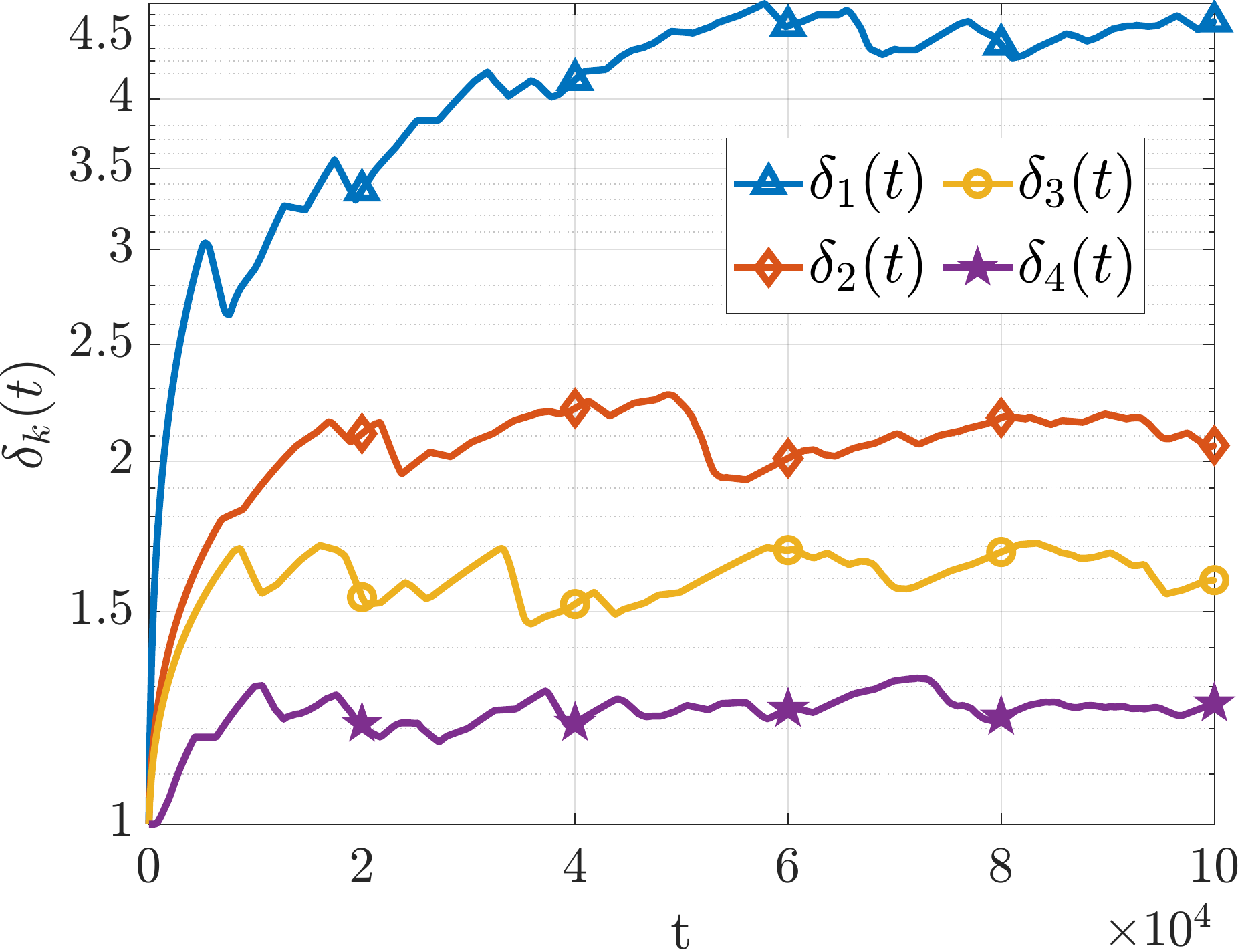}
    \label{fig:delta_adaptive}
    } 
     \subfloat[Out-of-service probability]{
    \includegraphics[width=0.32\textwidth]{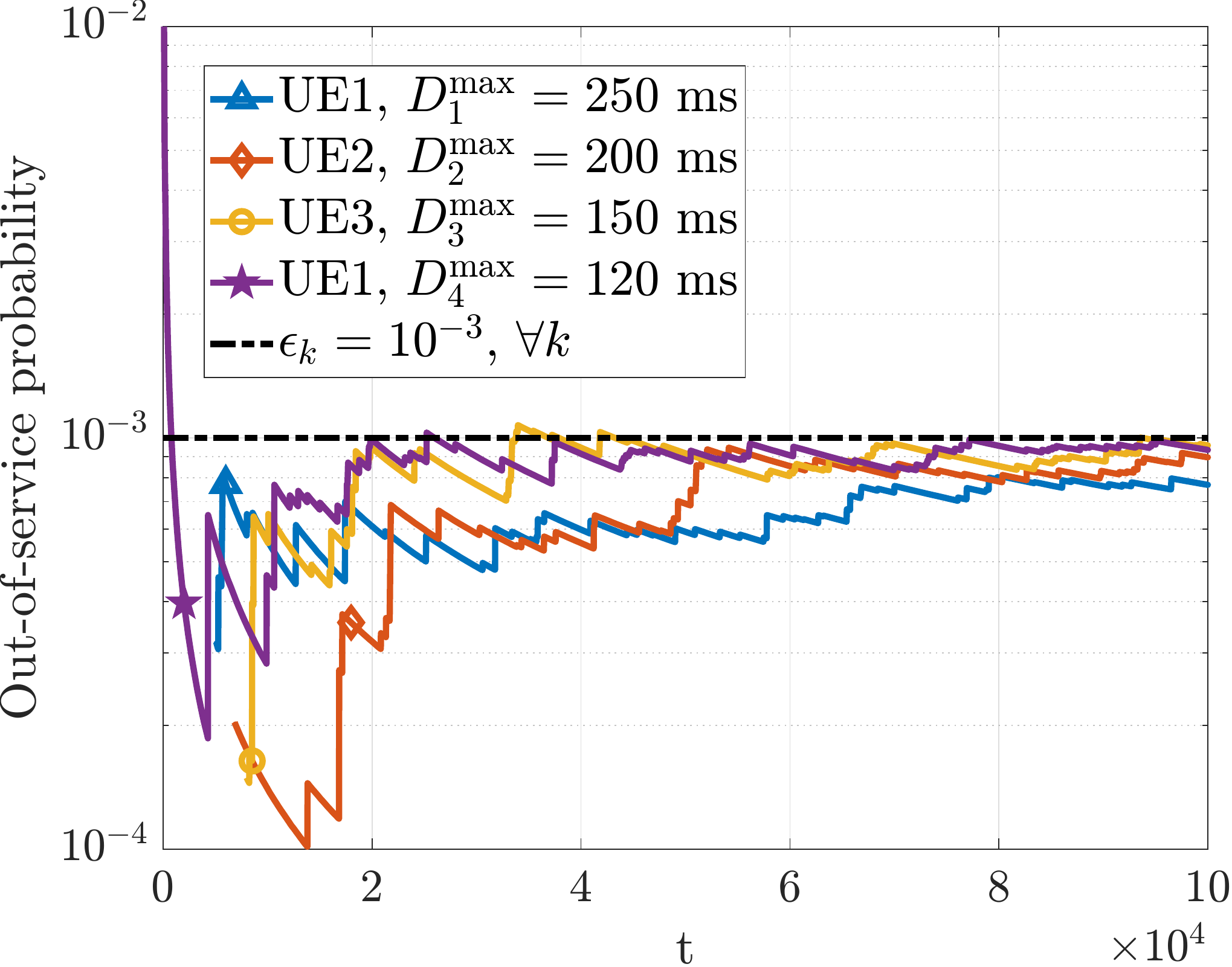}
    \label{fig:out_of_service_prob}
    } 
    \caption{Performance in terms of reliability}
\end{figure*}
\textit{\underline{Reliability:}}
Fig.~\ref{fig:reliability} focuses on the out-of-service constraint, i.e. constraint $(b)$ of \eqref{Problem}, 
and shows the effectiveness of the adaptive parameter $\delta_k$ in~\eqref{probabilistic}. The scenario is composed of $4$ UE, Poisson arrivals with $A_k^{\rm avg}=5$, $S_k^i=1000$ and $S_k^o=100$ bits, $J_k=10^{-4}$ data/CPU cycle, $D_k^{\rm avg}=100$ ms, $D_k^{\max}=[250,200,150,120]$ and a reliability requirement $\epsilon_k=10^{-3}$, with $\mu_k=20$. The adaptation of $\delta_k$ is obtained with starting point $\delta_k(0)=1$ $\forall k$, $\nu_k(0)=[15,5,4,3]$, $k=1,2,3,4$, and the diminishing rule in~\eqref{diminishing} uses $\beta_k=1/2$, $\forall k$. The probability of exceeding the desired maximum delay ($P_k(D_k^{\max},W_k^t)$ in \eqref{probs}) is estimated over the most recent $10^4$ data result arrivals (i.e. $|W_k(t)|=10^4$). The target PER is $10^{-4}$, and the trade-off parameter 
is $V=5\times 10^6$. The simulation is run for $10^5$ slots. Then, Fig.~\ref{fig:reliability} shows the reliability function (also known as survivor function), defined as $1-\text{CDF}(D_k)$, with $\text{CDF}(D_k)$ being the cumulative distribution function of the end-to-end delay experienced by all data of user $k$. The delay is measured by timestamping each data unit. Thus, each curve in Fig. \eqref{fig:reliability} shows the probability that the end-to-end delay of each data unit exceeds the value on the abscissa. The black dotted horizontal line represents the requirement $\epsilon_k$ on the out-of-service probability (cf. \eqref{probabilistic}). For each UE, the points corresponding to $D_k^{\max}$, $k=1,2,3,4$ are circled; they all lie below the horizontal black dotted line and the reliability constraint is met.
We also show, for each UE, the average energy consumption $E_k^{\rm avg} = \frac{1}{T} \sum_{t=1}^T \mathbb{E}\{E_k(t)\}$. In particular, 
the average system energy consumption resulting from the minimum delay strategy (i.e., always transmit) is $245$ mJ, while the average system energy consumption necessary to achieve the result of Fig. \ref{fig:reliability} is much lower ($160$ mJ). The evolution of $\delta_k(t)$ over time and its convergence are illustrated in Fig. \ref{fig:delta_adaptive}. 
As expected, a lower $D_k^{\max}$ requires a lower $\delta_k$. Finally, Fig. \ref{fig:out_of_service_prob} illustrates the instantaneous out-of-service probability obtained via the adaptive strategy, which flattens around 
$\epsilon_k$ after a transient interval. Note that the choice 
$\delta_k(0)=1$ is conservative and helps limiting the out-of-service probability when the convergence of the algorithm is not reached yet. Then, over time, the constraint is relaxed thanks to the adaptation rule of $\delta_k$, which helps reducing the energy consumption.\\
\textit{\underline{Comparison of different sleep modes strategies:}}
\begin{figure*}[t]
    \centering
    \subfloat[Avg. system energy vs $V$]{
    \includegraphics[width=0.318\textwidth]{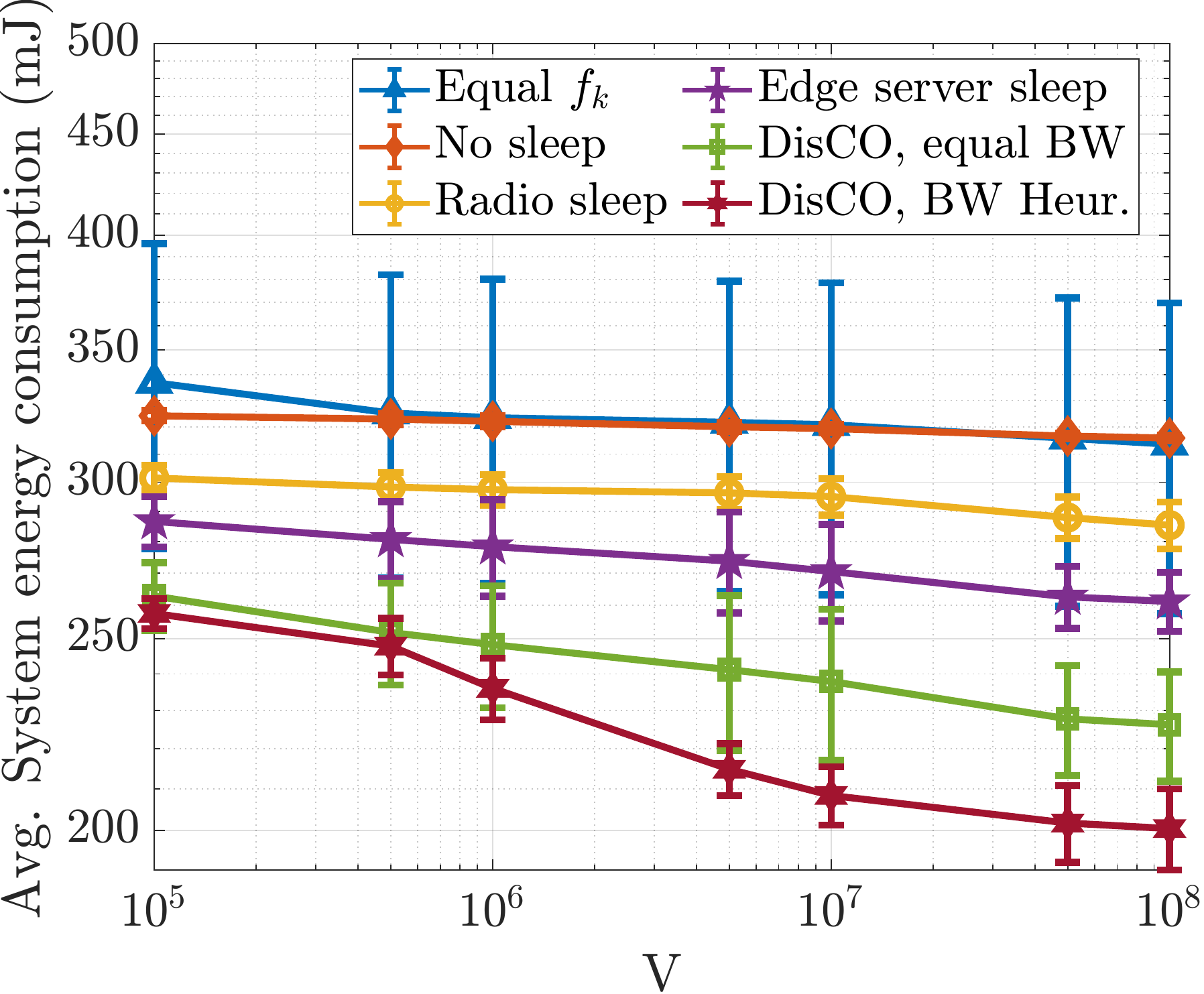}
    \label{fig:comparisons}
    }
    \subfloat[Avg. system energy vs. $A_k^{\rm avg}$]{
    \includegraphics[width=0.32\textwidth]{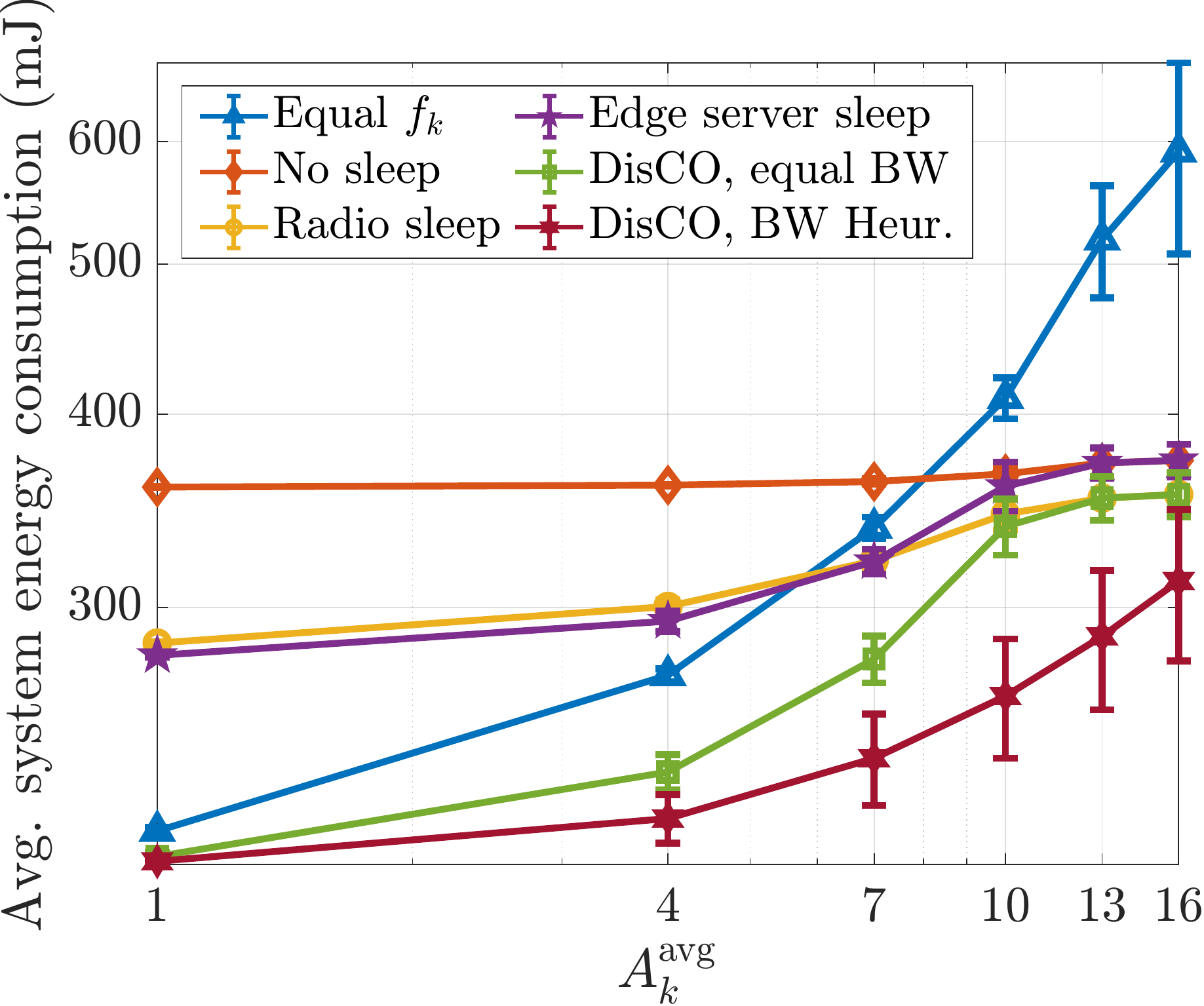}
    \label{fig:comparisons_arrivals}
    }
    \subfloat[Duty cycles vs. $A_k^{\rm avg}$]{
    \includegraphics[width=0.318\textwidth]{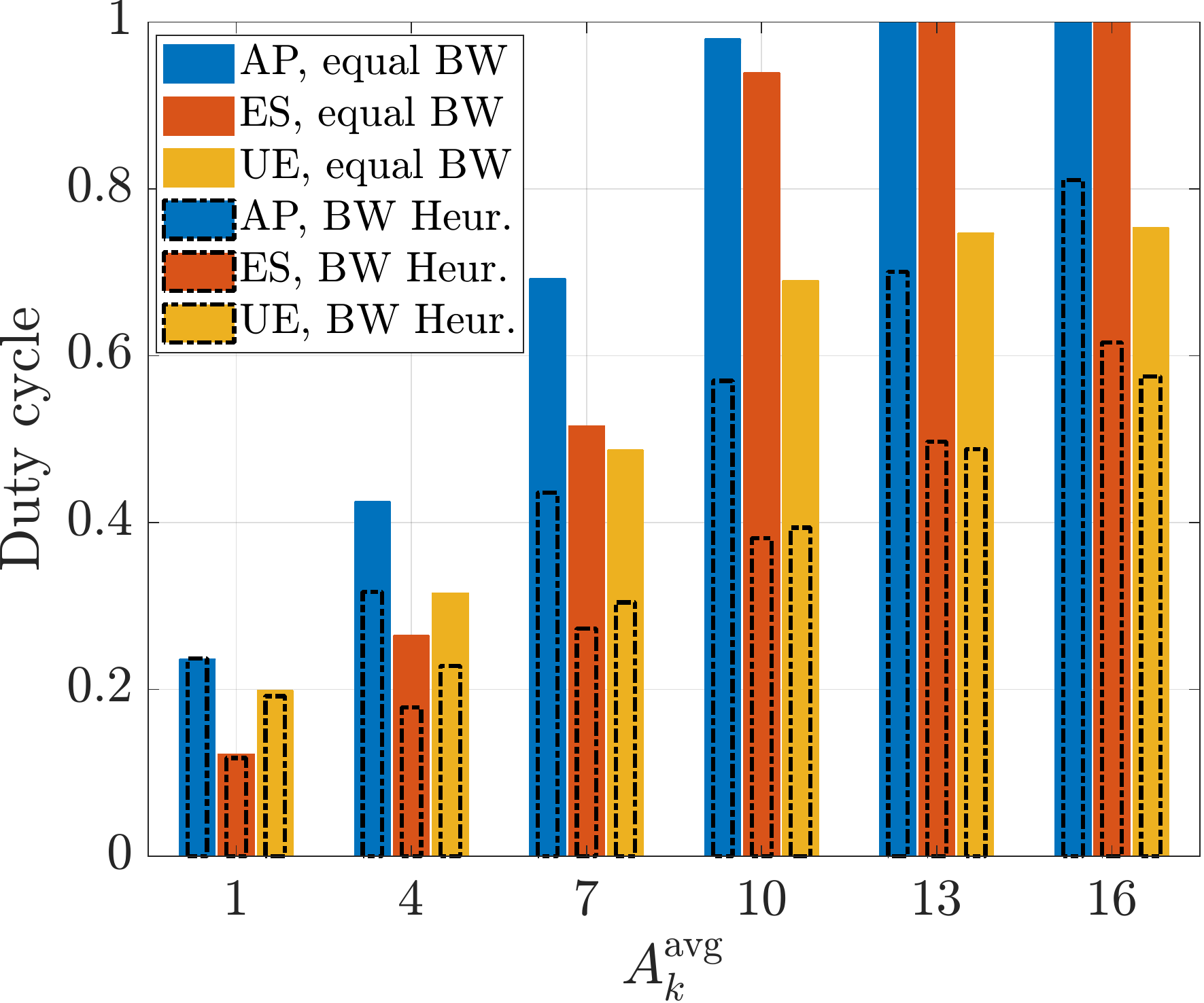}
    \label{fig:duties}
    }
    \caption{Comparison of DisCO with other strategies}
\end{figure*}
We now compare DisCO with four different resource strategies, which correspond to specific customization of other works to our setting. 
$i)$ \textit{Equal $f_k$'s}: resources are optimized (including $f_c$) but the CPU frequencies are equally allocated to each user by the ES (i.e. without Algorithm \ref{alg:cpu} for CPU scheduling).
$ii)$ \textit{No sleep}: resources are optimized but the network elements cannot be turned to sleep states. This could be possibly related to our previous work \cite{Merluzzi2020URLLC}, where we jointly optimize radio and computation resources in a user-centric fashion, not exploiting sleep modes.
$iii)$ \textit{Radio sleep}: resources are optimized but the sleep state of the ES is not available. This is analogous to the approach of \cite{Yu_Pu_2018}, once customized to our system model (i.e. single AP), where the authors only exploit AP sleep states.
$iv)$ \textit{ES sleep}: resources are optimized but the AP and the users cannot enter the sleep states. This is coherent with the results of \cite{Wang19}, where a sleep state at the ES is considered, but no sleep is exploited for UE and AP.
Also, we propose a different strategy for bandwidth allocation, based on the following heuristic: let 
$\widetilde{Q}_k^u=4Q_k^m-2Q_k^l+(Z_k+\mu_k Y_k)Q_k^l$ and 
$\mathcal{K}_u^+=\{k:\widetilde{Q}_k^u > 0\}$ for the uplink; 
similarly, let $\widetilde{Q}_k^d=4Q_k^a+(Z_k+\mu_k Y_k)Q_k^a$ and 
$\mathcal{K}_d^+=\{k:\widetilde{Q}_k^d>0\}$ for the downlink. We define the below uplink (downlink) bandwidth allocation rule:
\begin{align}\label{BW_u}
     &B_k^{u}=\begin{cases}
    \displaystyle \frac{\widetilde{Q}_k^{u}}{\sum_{i\in \mathcal{K}_{u}^+}\widetilde{Q}_i^{u}}B_{u}, \quad \textrm{if} \;\,k\in \mathcal{K}_{u}^+,\\
    0, \quad \textrm{otherwise},
    \end{cases} \nonumber\\
    &B_k^d=\begin{cases}
   \displaystyle \frac{\widetilde{Q}_k^d}{\sum_{i\in \mathcal{K}_d^+}\widetilde{Q}_i^d}B_d, \quad \textrm{if} \;\,k\in \mathcal{K}_d^+,\\
    0, \quad \textrm{otherwise},
    \end{cases}
\end{align}
where $B_u$ and $B_d$ are the total available uplink and downlink bandwidths, respectively. This heuristic for the allocation of spectral resources is based on the fact that all the information about the status of a certain UE's quality of service lies in the physical and virtual queues. Thus, a UE with a higher $\widetilde{Q}_k^u$ ($\widetilde{Q}_k^d$ for the downlink part), which is defined based on the objective function of \eqref{slot_opt_radio}, needs more resources to drain its queues.

We run our simulations with random configurations of the following parameters: $S_k^i=10^x$, $S_k^o=10^y$ bits, with $x$ and $y$ u.r.g. in $[1,3]$. We assume Poisson arrivals with $A_k^{\rm avg}$ u.r.g. in $[1,20]$ data units. Finally, $J_k=10^{-z}$ data/CPU cycle, with $z$ u.r.g. in $[2,5]$. We consider a scenario with $10$ users, all with an average delay requirement $D_k^{\rm avg}=[80,85,90,95,100,105,110,115,120,125]$ ms, and $\delta_k=[1.5,1.6,1.7,1.8,1.9,2.0,2.1,2.2,2.3,2.4]$. The simulation is run for $10^4$ slots and the results are averaged over $100$ independent realizations of the above parameters and UE' positions. In Fig. \ref{fig:comparisons}, we observe the non-negligible gain of DisCO in terms of average system energy consumption, when compared to all the proposed alternative strategies. The heuristic for bandwidth allocation described in \eqref{BW_u} 
and termed as ``DisCO (BW Heur.)'' in Fig.~\ref{fig:comparisons} achieves an additional gain around $10$\% with respect to DisCO with equal bandwidth allocation. Of course, other heuristics can be investigated and integrated with our strategy. For instance, at each $t$ (or a longer time scale), it is possible to compare the solutions obtained with different bandwidth allocation strategies and select the best one, if this is compatible with a practical implementation. A recent contribution suggests this possibility, with a parallel GPU based implementation \cite{Huang2018GPFAG}.\\
\textit{\underline{The effect of the arrival rate:}} In Fig.~\ref{fig:comparisons_arrivals}, we compare the average system energy consumption of DisCO with other strategies, considering different values of the parameter $A_k^{\rm avg}$, $\forall k$. 
The scenario involves $15$ UE; $S_k^i=10^x$, $S_k^o=10^y$ bits, with $x$ and $y$ u.r.g. in $[2,3]$ and $[1,3]$, respectively; $J_k=10^{-z}$ data/CPU cycle, with  $z$ u.r.g. in $[2,5]$; the average delay constraint is $D_k^{\rm avg}=100$ ms, $\delta_k=2$, $\epsilon_k=10^{-2}$, $\mu_k=10$, $\forall k$. 
The Lyapunov trade-off parameter is $V=5\times 10^7$. The simulation is run for $10^4$ slots and the results are averaged over $100$ independent realizations of the above parameters and UE positions. Fig.~\ref{fig:comparisons_arrivals} shows how DisCO is able to yield a large gain compared to the other strategies, except for high arrival rates, where there are less degrees of freedom to exploit the sleep mode operations. In particular, the duty cycles (fraction of activity time) obtained with DisCO are shown in Fig. \ref{fig:duties}, as a function of $A_k^{\rm avg}$. We considered the same setting used for Fig. \ref{fig:comparisons_arrivals}, using DisCO with equal bandwidth allocation, and with the heuristic described in \eqref{BW_u}. 
Fig. \ref{fig:duties} shows that, for high $A_k^{\rm avg}$, the duty cycles of DisCO are close to $1$ (i.e., always active), thus explaining the similar energy consumption as the strategies without sleep control. However, with our proposed heuristic for bandwidth allocation, we achieve a non-negligible gain in terms of activity time with respect to the equal bandwidth allocation strategy. This result further motivates taking into account the physical and virtual queues in prioritizing the scheduling of the users.
\section{Conclusions}
In this paper, we proposed a dynamic resource allocation algorithm for computation offloading that jointly exploits low-power sleep modes of UE, AP, and ES to reduce the system energy consumption with guaranteed E2E average delay and reliability. Via Lyapunov stochastic optimization, we solved a long-term problem, using a dynamic algorithm that works on a per-slot basis, without assuming any prior knowledge on the statistics of data arrivals and radio channels, and with theoretical guarantees. 
Several numerical results show the performance gain offered by our proposed online strategy, and how a holistic view of the system can be beneficial for all agents and for the global energy consumption. \textcolor{black}{In this paper, we focused on a multiuser setting with a single AP and single ES. Future investigations should include optimized scheduling of spectral and time radio resources in a multi-cell multi-server scenario, where the cooperation among multiple APs and ESs can help reducing the overall energy consumption. \textcolor{black}{Furthermore, non-cooperative methods, including purely game-theoretic approaches or incentive-based mechanisms (see e.g., \cite{ZhouChen2020,ZhouWu2021}), are worth of being  investigated, as a way to achieve distributed and efficient solutions, while minimizing signaling overhead. Finally, due to the partial knowledge of the communication and computation models involved, it is worth investigating both (partial) data-driven approaches, e.g. DRL methods, and (partial) model-based approaches, where whichever information, albeit limited, is incorporated and exploited to find efficient solutions.}}
\appendix
\section{appendix}
Here, we present the derivation of the upper bound of the Lyapunov drift-plus-penalty that leads to the per-slot optimization strategy in \eqref{slot_opt}. First of all, note that, given a generic virtual queue $X(t)$ evolving as
    $X(t+1)=\max(0,\ X(t)+x(t+1)-\bar{x})$, 
and defining $\Delta_X(t)=\frac{X^2(t+1)-X^2(t)}{2}$, we can always write $\Delta_X(t)\leq \frac{(x(t+1)-\bar{x})^2}{2}+X(t)x(t+1)-X(t)\bar{x}$ \cite[p. 59]{Neely10}.
Then, for the virtual queue $Z_k(t)$ defined in \eqref{virtual_Z}, we can write
\begin{align}\label{Z_first}
   & \Delta_Z(t)\leq \frac{\left(Q_k^{\rm tot}(t+1)-Q_k^{\rm avg}\right)^2}{2}+Z_k(t)Q_k^{\rm tot}(t+1)\nonumber\\
    &- Z_k(t)Q_k^{\rm avg}=\frac{1}{2}\left(Q_k^{\rm tot}(t+1)\right)^2+\frac{1}{2}\left(Q_k^{\rm avg}\right)^2\nonumber\\
    &-Q_k^{\rm tot}(t+1)Q_k^{\rm avg} +Z_k(t)Q_k^{\rm tot}(t+1)\!-\!Z_k(t)Q_k^{\rm avg}\nonumber\\
    &\leq \left(Q_k^l(t+1)\right)^2+ \left(Q_k^m(t+1)+Q_k^a(t+1)\right)^2+ \frac{1}{2}\left(Q_k^{\rm avg}\right)^2\nonumber\\
    &+ Z_k(t)Q_k^{\rm tot}(t+1) - Z_k(t)Q_k^{\rm avg}\leq \left(Q_k^l(t+1)\right)^2 \nonumber\\
    &+ 2\left(Q_k^m(t+1)\right)^2 + 2\left(Q_k^a(t+1)\right)^2+ \frac{1}{2}(Q_k^{\rm avg})^2 \nonumber\\
    &+ Z_k(t)Q_k^{\rm tot}(t+1)-Z_k(t)Q_k^{\rm avg}.
\end{align}
Now, for $A,b\geq0$ we have $(\max(0,Q-b)+A)^2\leq Q^2+A^2+b^2+2Q(A-b)$ \cite[p. 33]{Neely10}; recalling 
\eqref{q_loc}, \eqref{q_rem} and \eqref{q_ap} and applying the upper bound to all queues,
we can write
\begin{align}\label{upper_Z}
&\Delta_Z(t)\leq \left(Q_k^l(t)\right)^2 + (A_{k,\max})^2 +  (N_{k,\max}^{u})^2+ 2\left(Q_k^m(t)\right)^2\nonumber\\
    &+ 2Q_k^l(t)\left(A_k(t)-N_k^u(t)\right) + 2(N_{k,\max}^u)^2+ 2(N_{k,\max}^c)^2\nonumber\\
    & + 4Q_k^m(t)\left(N_k^u(t)-N_k^c(t)\right) +2\left(Q_k^a(t)\right)^2+ 2(N_{k,\max}^c)^2 \nonumber\\
    &+  2(N_{k,\max}^d)^2+ 4Q_k^a(t)\left(N_k^c(t) - N_k^d(t)\right) + \frac{1}{2}\left(Q_k^{\textrm{avg}}\right)^2\nonumber\\
    &+ Z_k(t)\big(\max\left(0,Q_k^l(t)-N_k^u(t)\right) + A_k(t)\nonumber\\
    &+\max \left(0,Q_k^m(t)-N_k^c(t)\right) + \min\left(Q_k^l(t),N_{k,\max}^u\right)\nonumber\\
    &+\max \left(0,Q_k^a(t)-N_k^d(t)\right)\nonumber\\
    &+\min\left(Q_k^m(t),N_{k,\max}^c\right)-Q_k^{\rm avg}\big),
\end{align}
where we used the fact that $\min(Q_k^m(t),N_k^c(t))\leq \min(Q_k^m(t),N_{k,\max}^c)$, and $\min\left(Q_k^l(t),N_k^u\right) \leq \min\left(Q_k^l(t),N_{k,\max}^u\right) $, where $N_{k,\max}^u$ is the maximum number of uplink transmitted data units, and $N_{k,\max}^c$ is the maximum number of computable data units, given \eqref{num_comp}.
For the virtual queue $Y_k(t)$ (cf. \eqref{virtual_Y}), we can write
\begin{align}\label{upper_Y}
    &\Delta_Y(t) \leq \frac{\mu_k^2 \left(u\{Q_k^{\textrm{tot}}(t+1)- \delta_k Q_k^{\rm avg}\}-\epsilon_k \right)^2}{2}\nonumber\\
    &+ \mu_k Y_k(t) \left(u\{Q_k^{\textrm{tot}}(t+1)-\delta_k Q_k^{\rm avg}\} - \epsilon_k \right)\nonumber\\
    &\leq \frac{\mu_k^2(1-\epsilon_k)^2}{2}+ \mu_kY_k(t) \left(u\{Q_k^{\textrm{tot}}(t+1)-\delta_k Q_k^{\rm avg}\} - \epsilon_k\right),
\end{align}
where we used the fact that $u\{\cdot\}\leq 1$. Finally, plugging \eqref{upper_Z} and \eqref{upper_Y} into 
\eqref{drift_plus_penalty}, we can write 
\begin{align}\label{upper_bound}
&\Delta_p(\mathbf{\Theta}(t)) \leq \zeta +\mathbb{E} \bigg\{ \sum_{k=1}^K \bigg[\chi_k(t)-2Q_k^l(t)N_k^u(t)\nonumber\\
&+ 4Q_k^m(t)\left(N_k^u(t) - N_k^c(t)\right)+ 4Q_k^a(t) \left(N_k^c(t) - N_k^d(t)\right)\nonumber\\
&+ Z_k(t)\big(\max\left(0,Q_k^l(t)-N_k^u(t)\right)\nonumber\\
&+\max\left(0,Q_k^m(t)-N_k^c(t)\right)+\max\left(0,Q_k^a(t)-N_k^d(t)\right)\big)\nonumber\\
&+ \mu_k Y_k(t)u\bigg\{\max\left(0,Q_k^l(t)-N_k^u(t)\right)+A_k(t)\nonumber\\
&+\max\left(0,Q_k^m(t)-N_k^c(t)\right)+\min(Q_k^l(t),N_{k,\max}^u(t))\nonumber\\
&+\max\left(0,Q_k^a(t)-N_k^d(t)\right)+\min(Q_k^m(t),N_{k,\max}^c(t))\nonumber\\
&-\delta_k Q_k^{\rm avg}\bigg\}\bigg]+ V E_{\rm tot}^w(t)\bigg|\mathbf{\Theta}(t)\bigg\},
\end{align}
where $\zeta$ is a positive constant given by
\begin{align}\label{zeta}
    &\zeta = \sum_{k=1}^K \big[\left(A_{k,\max}\right)^2 + 3\left(N_{k,\max}^{u}\right)^2 + 4\left(N_{k,\max}^c\right)^2\nonumber\\
    &+ 2\left(N_{k,\max}^d\right)^2\!+\! \frac{\left(Q_k^{\textrm{avg}}\right)^2}{2} \!+\! \frac{\mu_k^2(1-\epsilon_k)^2}{2}\big],
\end{align}
and $\chi_k(t)$ is a constant at time slot $t$ (i.e. it does not depend on the optimization variables), which reads as follows:
\begin{align}\label{chi}
&\chi_k(t)=(2Q_k^l(t)+Z_k(t))A_k(t)+(Q_k^l(t))^2+2(Q_k^m(t))^2\nonumber\\
&+2\left(Q_k^a(t)\right)^2+Z_k(t)(\min(Q_k^l(t),N_{k,\max}^u(t))\nonumber\\
&+\min(Q_k^m(t),N_{k,\max}^c(t))-Q_k^{\rm avg})-\mu_kY_k(t)\epsilon_k.
\end{align}
Then, the Min-Drift-plus penalty algorithm proceeds by opportunistically minimizing \eqref{upper_bound} in each time slot, leading to the problem in \eqref{slot_opt}, where all the constant terms (with respect to the variables) are omitted.

\bibliographystyle{IEEEtran}
\bibliography{IEEEabrv,Mattia}
\begin{IEEEbiography}[{\includegraphics[width=1 in,height=1.25in,clip,keepaspectratio]{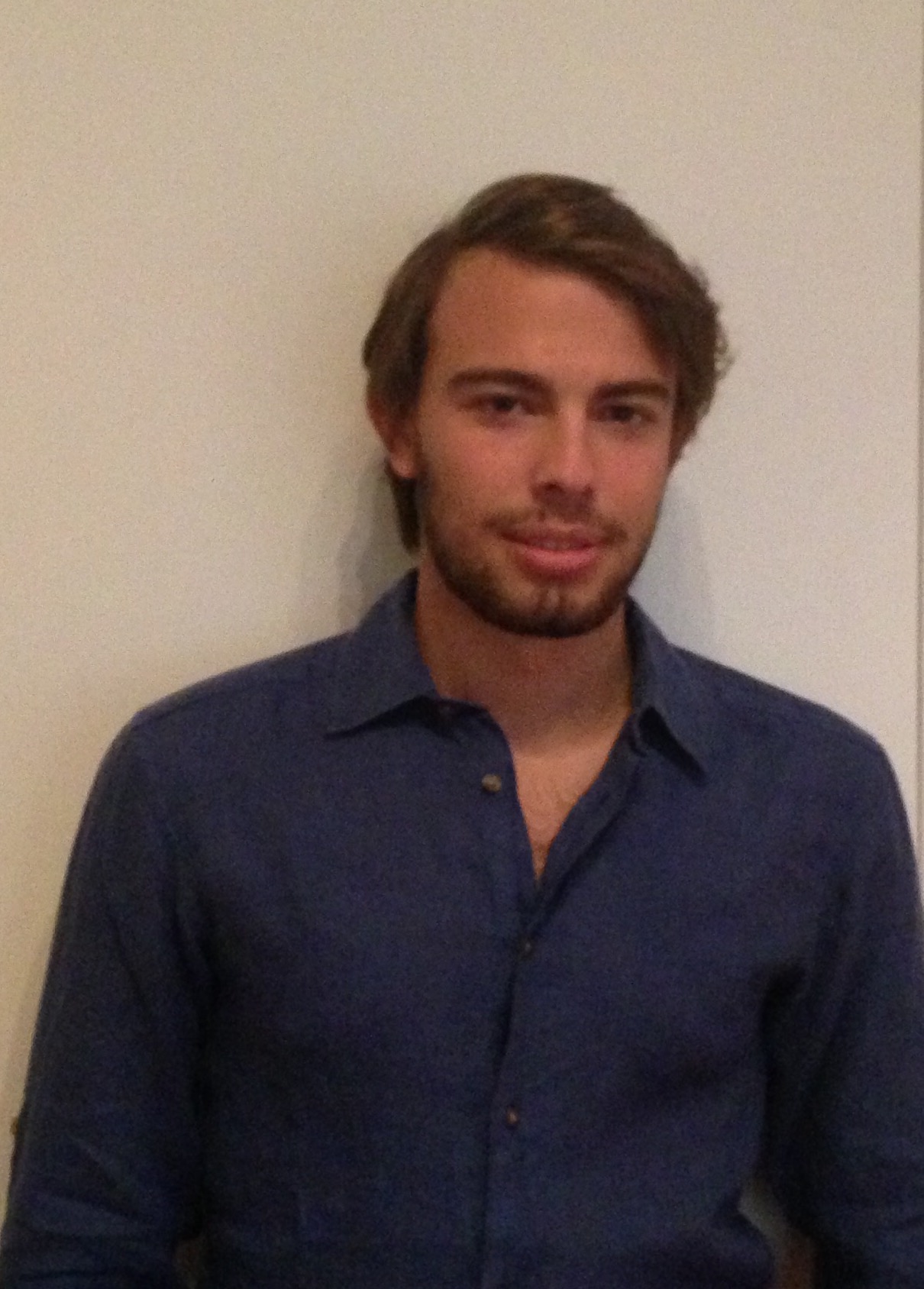}}]
{Mattia Merluzzi} (Member, IEEE) received the M.S. degree in Telecommunication Engineering and the Ph.D. degree in Information and Communication Technologies from Sapienza University of Rome, Italy, in 2017 and 2021, respectively. He is currently a research engineer at CEA-Leti, Grenoble, France, where he is involved in the research team of the H2020 project Hexa-X. He has participated in the H2020 EU/Japan project 5G-Miedge, the H2020 EU/Taiwan project 5G CONNI and the MIUR funded PRIN Liquid Edge. His primary research interests are in edge computing, beyond 5G systems, stochastic optimization, and edge machine learning. He was the recipient of the 2021 GTTI (Italian National Group on Telecommunications and Information Theory) Award for the Best Ph.D. thesis.
\end{IEEEbiography}
\vspace{-1 cm}
\begin{IEEEbiography}[{\includegraphics[width=1in,height=1.25in,clip,keepaspectratio]{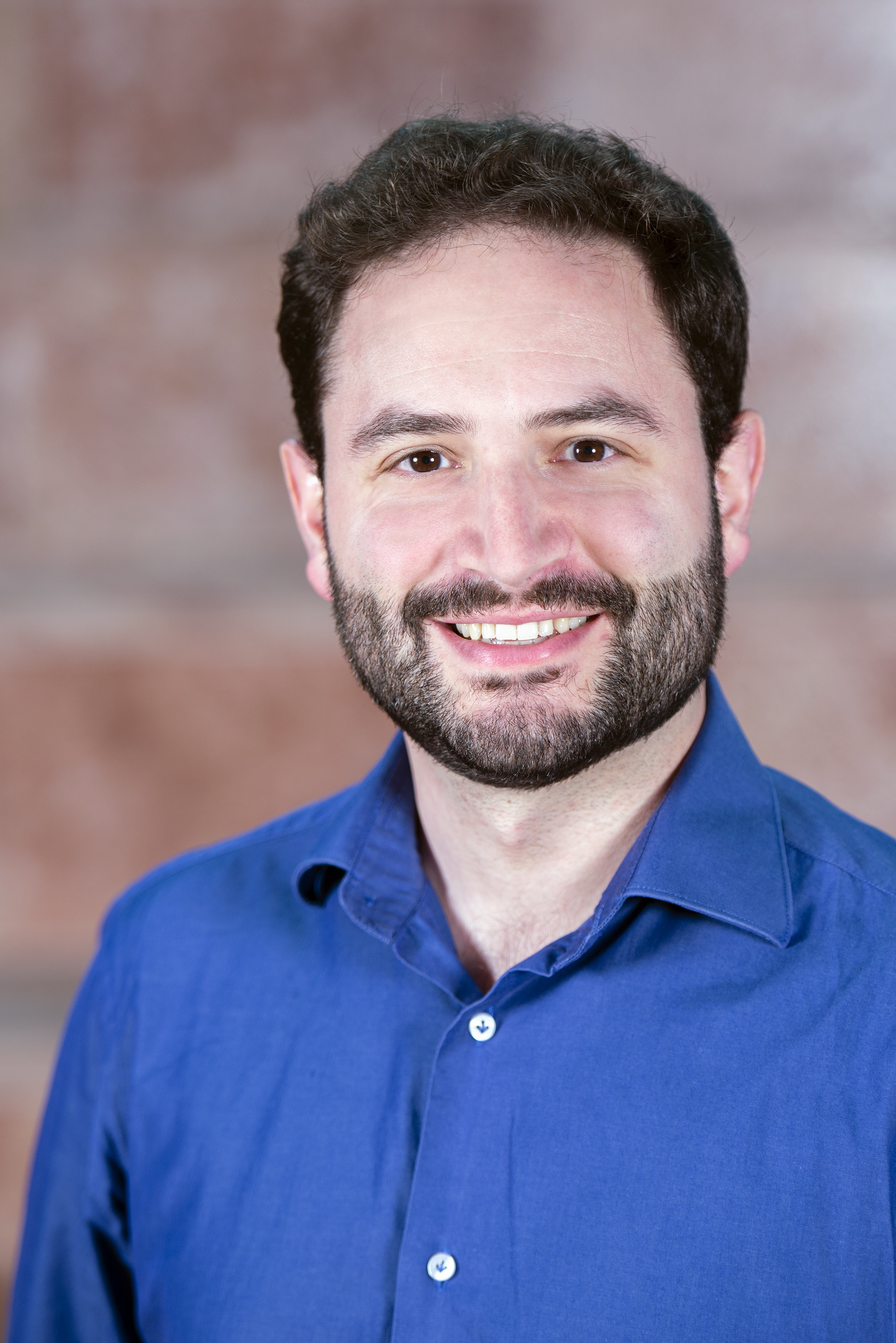}}]{Nicola di Pietro}
received the B.S. degree in mathematics from the University of Padova, Italy, in 2008. In 2010, he received the M.S. degree in mathematics jointly from the University of Padova, Italy, and the University of Bordeaux, France, within the framework of the international ALGANT program. He received the Ph.D. degree in mathematics from the University of Bordeaux, France, in 2014. During the years of his doctoral studies, he was a Research Engineer with the European R\&D Center of Mitsubishi Electric in Rennes, France. From 2014 to 2016, he was an Associate Post-Doctoral Fellow at Texas A\&M University at Qatar. From 2017 to 2021, he was a Research Engineer with CEA-Leti in Grenoble, France. He is now a System Engineer at Athonet, Italy. He is author of several papers and patents, and his research interests are 5G networks, edge computing, information theory, and lattice error-correcting codes.
\end{IEEEbiography}
\vspace{-1 cm}
\begin{IEEEbiography}[{\includegraphics[width=1in,height=1.25in,clip,keepaspectratio]{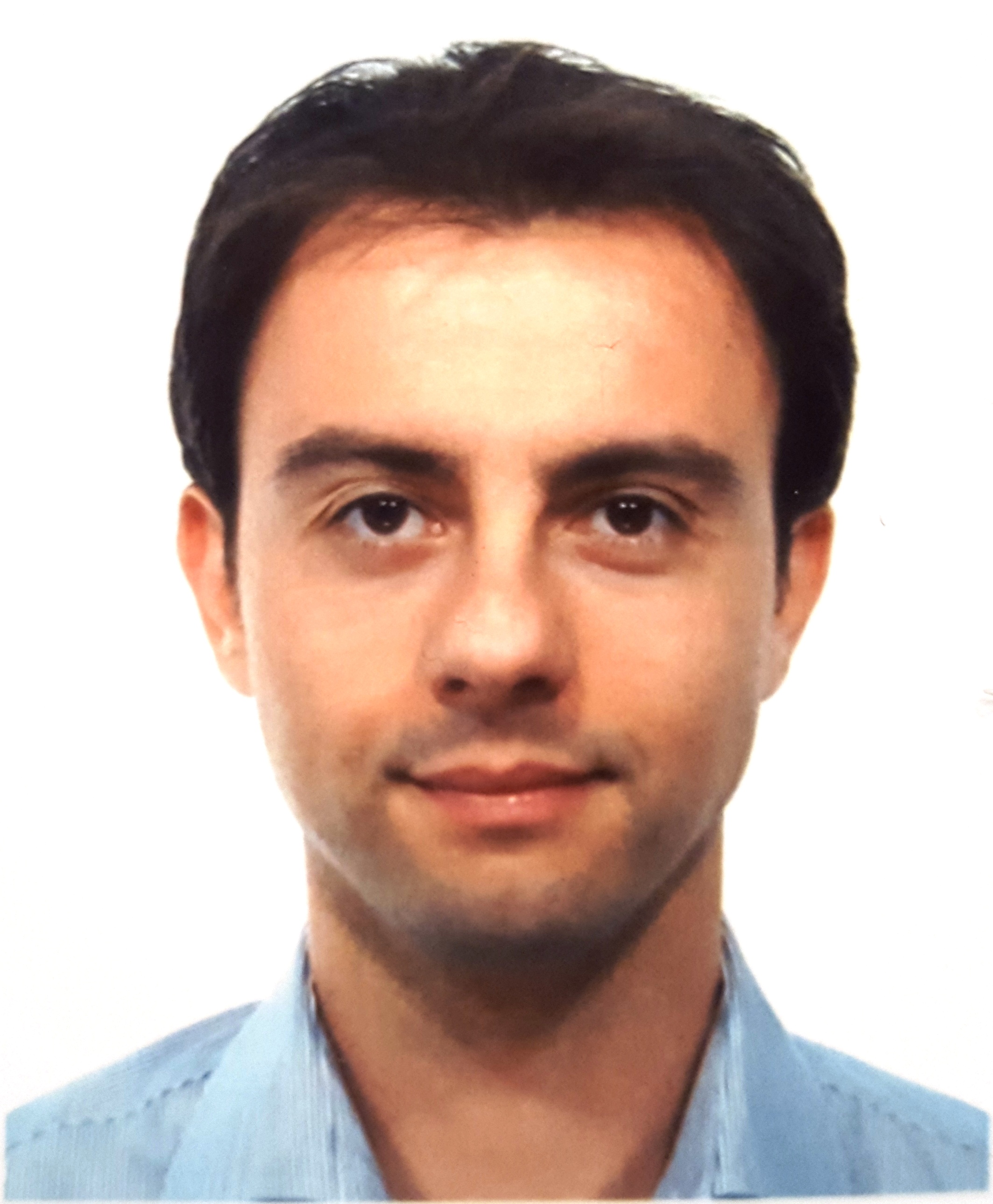}}] {Paolo Di Lorenzo} (Senior Member, IEEE) received the M.Sc. and the Ph.D. degrees in electrical engineering from Sapienza University of Rome, Rome, Italy, in 2008 and 2012, respectively. He is currently an Associate Professor with the Department of Information Engineering, Electronics, and Telecommunications, Sapienza University of Rome. In 2010, he held a visiting research appointment with the Department of Electrical Engineering,
University of California at Los Angeles, Los Angeles, CA, USA. From May 2015 to February 2018, he was an Assistant Professor with the Department of Engineering, University
of Perugia, Perugia, Italy. He has participated in the FP7 European research
projects FREEDOM, on femtocell networks; SIMTISYS, on moving target detection and imaging using a constellation of satellites; and TROPIC, on communication, computation, and storage over collaborative femtocells. He is a Principal Investigator of the research unit in the H2020 European project
RISE 6G. His research interests include signal processing theory and methods,
distributed optimization, mobile edge computing, machine learning, and graph
signal processing. Prof. Di Lorenzo is currently an Associate Editor for the IEEE Transactions on Signal and Information Processing Over Networks. He was the recipient of the three best student paper awards, respectively, at IEEE SPAWC10, EURASIP EUSIPCO11, and IEEE CAMSAP11. He was also the recipient of the 2012 GTTI (Italian National Group on Telecommunications and Information Theory) Award for the Best Ph.D. thesis.
\end{IEEEbiography}
\begin{IEEEbiography}[{\includegraphics[width=1in,height=1.25in,clip,keepaspectratio]{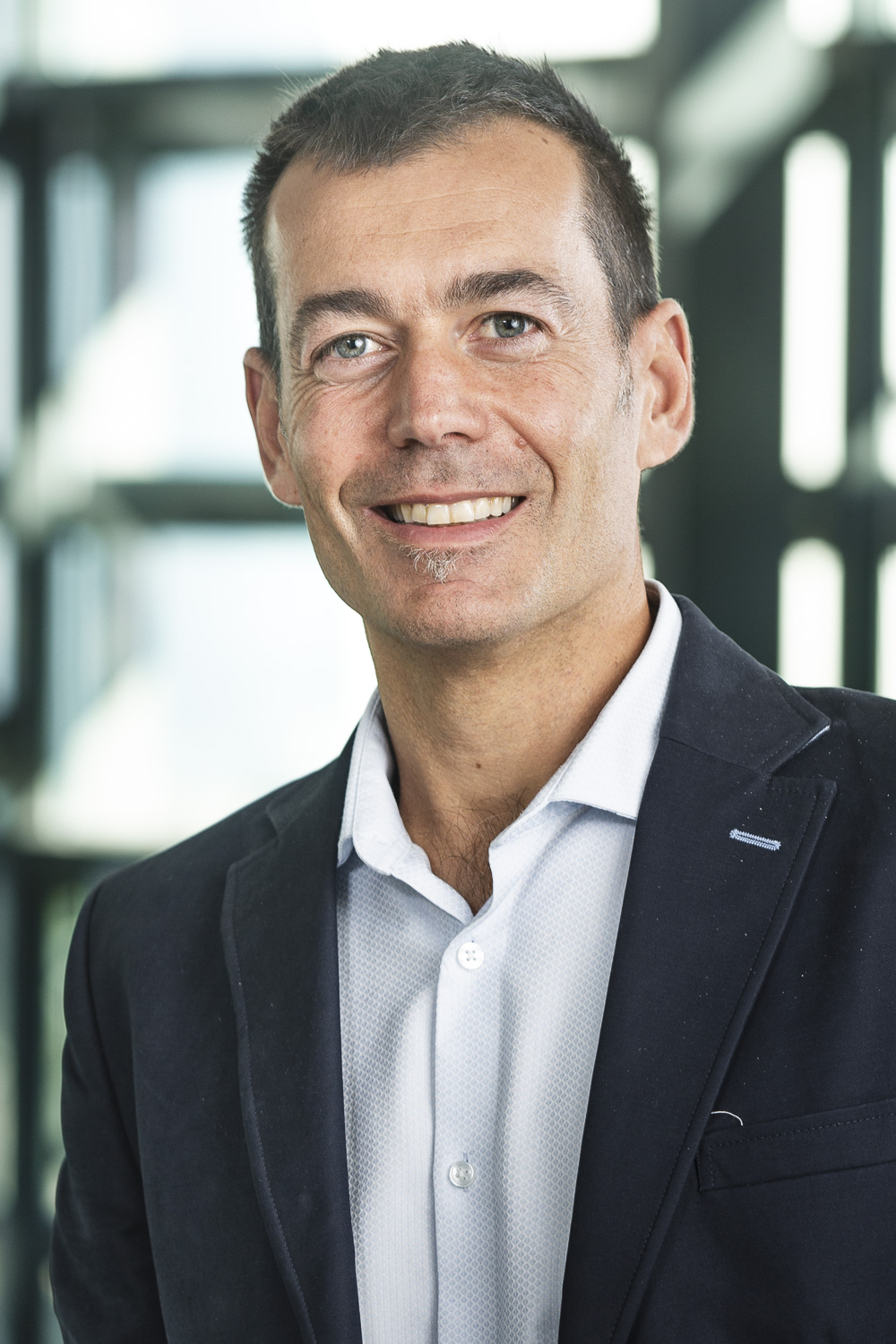}}]{Emilio Calvanese Strinati} (Member, IEEE) obtained his Engineering Master degree in 2001 from Sapienza University of Rome and his Ph.D in Engineering Science in 2005. He then started working at Motorola Labs in Paris in 2002. Then in 2006 he joined CEA-Leti as a research engineer. From 2007, he becomes a PhD supervisor. From 2010 to 2012, he has been the co-chair of the wireless working group in GreenTouch Initiative which deals with design of future energy efficient communication networks. From 2011 to 2016 he was the Smart Devices \& Telecommunications European collaborative strategic programs Director. Between December 2016 and January 2020 is was the Smart Devices \& Telecommunications Scientific and Innovation Director. From 2017 to 2018 he was one of the three moderators of the 5G future network expert group. Between 2016 and 2018 he was the coordinator of the H2020 joint Europe and South Korea 5GCHAMPION project that showcased at the 2018 winter Olympic Games, 5G technologies in realistic operational environments. Since July 2018 he is the coordinator of the H2020 joint Europe and South Korea 5G-AllStar project. Since 2018 he holds the French Research Director Habilitation (HDR). In 2021 he started the coordination of the H2020 European project RISE-6G, focusing on the design and operation of Reconfigurable Intelligent Surfaces in future high frequency 6G networks. Since February 2021 he is also the director of the New-6G (Nano Electronic \& Wireless for 6G) initiative , dedicated to the required convergence between microelectronic \& telecom, hardware \& software, network \& equipment for upcoming 6G technologies.
He has published around 150 papers in international conferences, journals and books chapters, given more than 200 international invited talks, keynotes and tutorials. He is the main inventor or co-inventor of more than 70 patents. He has organized more than 100 international conferences, workshops, panels and special sessions on green communications, heterogeneous networks and cloud computing hosted in international conferences as IEEE GLOBCOM, IEEE PIMRC, IEEE WCNC, IEEE ICC, IEEE VTC, EuCNC, IFIP, EuCNC and European Wireless. He is the general chair of EuCNC 2022.

\end{IEEEbiography}

\begin{IEEEbiography}[{\includegraphics[width=1in,height=1.25in,clip,keepaspectratio]{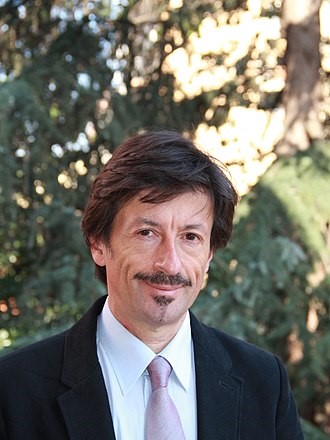}}]
{Sergio Barbarossa} (Fellow, IEEE) received his MS and Ph.D. EE degree from the Sapienza University of Rome, where he is now a Full Professor and Senior Research Fellow of the Sapienza School of Advanced Studies. He has held visiting positions at the Environmental Research Institute of Michigan ('88), Univ. of Virginia ('95, '97), and Univ. of Minnesota ('99). He is an IEEE Fellow, EURASIP Fellow, and he has been an IEEE Distinguished Lecturer. He received the IEEE Best Paper Award from the IEEE Signal Processing Society in the years 2000, 2014, and 2020. He received the Technical Achievements Award from the EURASIP Society in 2010. He  coauthored the papers that received the Best Student Paper Award at ICASSP 2006, SPAWC
2010, EUSIPCO 2011, and CAMSAP 2011. He has been the scientific coordinator of several EU projects on wireless sensor networks, small cell networks, distributed mobile cloud computing, and edge computing in 5G networks. He is now leading a national project on edge learning and he is involved in two H2020 European projects on 5G networks for Industry 4.0 and on reconfigurable intelligent surfaces. His current research interests are in the area of mobile edge computing and machine learning, graph signal processing, and distributed optimization. From 1997 to 2003, he was a member of the IEEE
Technical Committee for Signal Processing in Communications. He served as
an Associate Editor for the IEEE TRANSACTIONS ON SIGNAL PROCESSING
(1998-2000 and 2004-2006), the IEEE SIGNAL PROCESSING MAGAZINE, and the IEEE TRANSACTIONS ON SIGNAL AND INFORMATION PROCESSING OVER NETWORKS. He has been the
General Chairman of the IEEE Workshop on Signal Processing Advances
in Wireless Communications (SPAWC), 2003 and the Technical Co-Chair of
SPAWC, 2013. He has been the Guest Editor for Special Issues on the IEEE
JOURNAL ON SELECTED AREAS IN COMMUNICATIONS, EURASIP Journal of Applied Signal Processing, EURASIP Journal on Wireless Communications and Networking, the IEEE SIGNAL PROCESSING MAGAZINE, and the IEEE SELECTED TOPICS ON SIGNAL PROCESSING.
\end{IEEEbiography}
\end{document}